\documentclass[11pt]{article}
\usepackage[bottom=2.5cm,top=3.0cm,left=3cm,right=3cm]{geometry}
\usepackage[utf8]{inputenc}
\RequirePackage[OT1]{fontenc}
\RequirePackage{amsthm,amsmath,graphicx,float}
\usepackage[flushleft]{threeparttable}
\RequirePackage[colorlinks,citecolor=blue,urlcolor=blue]{hyperref}
\usepackage{caption}
\usepackage{subcaption}
\usepackage{booktabs}
\usepackage{multirow}

\title{Bayesian spatio-temporal model with INLA for dengue fever risk prediction in Costa Rica}
\author{Shu-Wei Chou-Chen \thanks{corresponding author: shuwei.chou@ucr.ac.cr}, Luis A. Barboza, Paola V\'asquez, Yury E. Garc\'ia, \\Juan G. Calvo, Hugo G. Hidalgo, Fabio Sanchez}

\begin{document}

\maketitle

\begin{abstract}
Due to the rapid geographic spread of the \emph{Aedes} mosquito and the increase in dengue incidence, dengue fever has been an increasing concern for public health authorities in tropical and subtropical countries worldwide. Significant challenges such as climate change, the burden on health systems, and the rise of insecticide resistance highlight the need to introduce new and cost-effective tools for developing public health interventions. Various and locally adapted statistical methods for developing climate-based early warning systems have increasingly been an area of interest and research worldwide. Costa Rica, a country with micro-climates and endemic circulation of the dengue virus (DENV) since 1993, provides ideal conditions for developing projection models with the potential to help guide public health efforts and interventions to control and monitor future dengue outbreaks.  

\end{abstract}



\section{Introduction}


Dengue is one of the most prevalent vector-borne diseases globally, affecting individuals of all ages. The infection can be asymptomatic or cause a broad spectrum of clinical manifestations that range from a non-specific and auto-limited viral syndrome to a disease with hemorrhagic manifestations and multi-systemic damage that can lead to the death of the patient~\cite{WHO:online}. The infection is caused by one of four dengue virus serotypes (DENV 1–4) transmitted to humans through the bite of infected female mosquitoes, primarily by \textit{Aedes aegypti} and \textit{Aedes albopictus} as a secondary vector. 

The interaction of a variety of factors, including globalization, trade, travel, demographic trends, and warming temperatures, have been associated with the spread of the mosquito, which is now present on all continents except Antarctica~\cite{medlock2006analysis,romi2006cold}, making it one of the 100 worst invasive species in the world~\cite{outammassine2022global}. It has also led to the emergence of the disease in places where it was previously absent~\cite{murray2013epidemiology,massad2018estimating,lopez2016modeling,gubler2012economic,wang2022global}, putting more than half of the world's population at risk of infection, mainly in tropical and subtropical regions~\cite{WHO:online}. In this context, and without effective prevention and control measures, dengue is expected to continue its geographical expansion~\cite{messina2019current,yang2021global}. Due to the lack of a vaccine and antiviral drugs, the key to preventing and controlling outbreaks continues to be the reduction of breeding sites through chemical and biological interventions as well as the active involvement of the community, which is mainly dependent on available resources of the affected countries. 

Given that the ecology of the virus is intrinsically tied to the ecology of mosquitoes that transmit dengue~\cite{morin2013climate}, climatic and environmental conditions can alter spatial and temporal dynamics of vector ecology, especially temperature, rainfall, and relative humidity~\cite{naish2014climate,campbell2015weather}. Temperature affects viral amplification~\cite{watts1986effect}, increases vector survival, reproduction, and biting rate~\cite{tun2000effects,rueda1990temperature}. Long-term breeding habitats for eggs, larvae, and pupae may be influenced by wastewater left by rainfall or human behavior of storing water in  containers~\cite{sarfraz2012analyzing,dieng2012household}. Dengue incidence has also been associated with vegetation indices, tree cover, housing quality, and surrounding land cover~\cite{van2005spatial,barrera2006ecological}.


Understanding the influence of these climatic variables on disease incidence in different regions can lead to early detection of disease progression, guide resource allocation, and implement appropriate health intervention~\cite{WHO:online}. In this effort, several methods of surveillance systems have been developed~\cite{mateus2011predictors,eidson2001dead, VBORNET:online}. However, successful early warning strategies are limited due to the complex and dynamic nature of the disease. The complex interaction of biological, socioeconomic, environmental, and climatic factors creates substantial spatio-temporal heterogeneity in the intensity of dengue. It imposes a challenge in the creation of surveillance and control systems.

In Costa Rica, a Central American country with a variety of micro-climates in an area of 51,179 $km^2$, dengue has been endemic since 1993 and has represented a public health burden since then. According to the Ministry of Health, more than 398,000 cases have been reported during the last 28 years~\cite{SitioWeb58:online}. DENV-1, DENV-2, and DENV-3 have been the main serotypes in circulation. However, it is essential to highlight that in 2022 the circulation of serotype four was identified in different municipalities in the country. Since this serotype has historically been absent in national territory, its emergence and lack of immunity in the country which can lead to an increase in the incidence of cases reinforces the need to increase prevention, detection, and timely treatment efforts and thus avoid an increase in the incidence and evolution of more severe forms of the disease. 

In this work, we propose a spatio-temporal model to predict dengue in Costa Rica, including climatic variables and geographic information to capture the effect of factors that modulate the spatio-temporal variation of dengue incidence in the country and whose information is not available, such as population mobility, socioeconomic and demographic information. This study is part of a series of efforts~\cite{vazquez2020,garcia2023wavelet}, which have been carried out to develop an early warning system to monitor dengue risk in the country. The aim is to provide guidance tools to the health authorities of Costa Rica that can be implemented and validated and to optimize and distribute resources in the prevention and control of dengue.

Based on dengue's historical incidence burden and suitable environment for disease transmission, the Costa Rican health authorities have identified 32 municipalities of interest (out of the 83 municipalities the country is divided) where the study was conducted. The selection included mainly municipalities located in the coastal areas on the Pacific and Caribbean coasts and municipalities in the Great Metropolitan Area, the country's most urban and populated region. This selection was also based on the decision-makers necessity to include new and cost-effective tools to guide the allocation of resources throughout the year. The article is divided as follows: Section 2 describes the data, statistical model, assumptions, and implemented methodologies. Section 3 presents the results for the 32 municipalities of interest. Finally, section 4 discusses this modeling approach's results, limitations, advantages, and future work.


\section{Methods}

\subsection{Data description}
\subsubsection{Dengue Cases}

For this analysis, we used monthly dengue cases for 32 municipalities of interest to public health authorities in Costa Rica. The data covered 2000-2021 obtained from the Ministry of Health of Costa Rica~\cite{minsa:online}.

\subsubsection{Climate variables}

\begin{enumerate}
\item Daily Precipitation estimates ($P_{i,t}$) were used to index land surface rainfall. Data were obtained from the Climate Hazards Group InfraRed Precipitation with Station data (CHIRPs); see~\cite{FunkChris2015Tchi}. Due to the high-resolution spatial nature of this dataset (5km by 5km), we were able to compute monthly cumulative rainfall estimates for each municipality by adding the exact estimate over smaller administrative areas (\textit{distritos}). 
	
\item El Ni\~no Southern Oscillation (ENSO, $S_{i,t}$) variations were indexed using the Sea Surface Temperature Anomaly (SSTA) index for the region known as Niño 3.4 (5N-5S, 120W-170W). monthly data was obtained from the Climate Prediction Center (CPC) of the United States National Oceanographic and Atmospheric Administration (NOAA) (see~\cite{CPC-SSTA}).
	
\item Normalized Difference Vegetation Index (NDVI, $N_{i,t}$), an index of the greenness of vegetation for a 16-day time resolution and 250m spatial resolution. It was obtained from the Moderate Resolution Imaging Spectroradiometer (MODIS) satellite and available through the \verb|MODISTools| R package (see~\cite{TuckPhillips}).
	
\item Daytime Land Surface Temperature (LST, $L_{i,t}$) in degrees Kelvin for an 8-day time resolution and 1km spatial resolution were obtained using the same resources as the NDVI covariate.
	
\item Tropical Northern Atlantic Index (TNA, $TN_{i,t}$). Anomaly index of the Sea-Surface Temperature Anomaly (SSTA) over the eastern tropical North Atlantic Ocean (see~\cite{EnfieldDavidB1999Huit}). Previous work in the region~\cite{Hidalgo2017-hr} suggested that including SSTA information from the Caribbean/Atlantic improves the performance of the prediction of land surface precipitation and temperature in Central America compared to forecasts produced with only Pacific Ocean ENSO conditions.   
\end{enumerate}

\subsection{Model}

We incorporate the historical exposure of the climate covariates and the behavior of the relative risks in the past by applying the Distributed Lag Non-Linear Model (DLNM) framework ~\cite{Gasparrini2010,Gasparrini2014}. This methodology incorporates a bi-dimensional space of functions that specifies an exposure-lag-response function $f \cdot w (x,l)$, which depends on the predictor $x$ along the time lags $l$ in a combined way. This combination specifies a non-linear and delayed association between a climate covariate and dengue incidence. For each covariate, we consider a minimum of 3-month lag and a maximum exposure of 12 months to obtain the estimates up to a maximum of a three-month ahead prediction. This combination of lags was determined based on the cross-correlation and wavelet behavior among the series (see \cite{garcia2023wavelet}). We also tested the model with different combinations of maximum and minimum lags. 

To compare the predictive gain of the non-linear specification against the linear ones, we compared models with a non-linear relationship of the delayed effect of exposure on the outcome using natural cubic B-splines function with 2 knots for each covariate and delayed effects against models with the linear relationship of the delayed effect of exposure on the outcome. 

After specifying the historical exposure of the covariates, we use a spatio-temporal Bayesian hierarchical model with the response variable as the monthly number of cases of dengue fever for each municipality $i, i=1,...,32$ for $t=1,...,264$ as follows:
\begin{align}
    \label{eq:model}
	Y_{it} | \mu_{it}, \kappa \sim NegBin (\mu_{it}, \kappa), \\
	\log(\mu_{it})= \log(E_{it})+ \log (RR_{it}),
\end{align}
and
\begin{multline*}
	\log RR_{it}=\alpha + f_1(RR_{t})+f_2(P_t)+ f_3(S_t)+ f_4(N_t)\\
	+ f_5(L_t)+f_6(TN_t)+ f_7(M_t)+ \phi_{i,(month)} + \theta_{i,(year)}, 
\end{multline*}
where $f_k,k=1,...,7$ is the exposure-lag-response function that applies a linear effect on each climate covariate from lag $3$ to $12$; $\phi_{i,(month)}$ is the municipality-specific monthly random effect that follows a prior according to a cyclic random walk of order 1, i.e., $\phi_{i,(month)}-\phi_{i,(month-1)} \sim N(0,\sigma^2_\phi)$; and $\theta_{i,(year)}$ is a random spatial effect.

For the spatial effect, two types of proximity matrix $\boldsymbol{W}$ are defined:
\begin{enumerate}
    \item The usual neighbor matrix is defined by $\boldsymbol{W}= \boldsymbol{\left\lbrace W\right\rbrace}_{ij} = 1$ if municipality $i$ and $j$ are neighbors, and $0$ otherwise.
    \item An alternative distance matrix based on the main road distance in kilometers between the central downtown of each pair of municipalities, i.e.
    $\boldsymbol{W} = \boldsymbol{\left\lbrace W\right\rbrace_{ij}} = 1$ if the distance is less than its overall median and $0$ otherwise. We incorporate this distance to provide a more realistic way to measure the proximity between social dynamics.
\end{enumerate}

Four types of spatial structures are implemented. First, the independent case is assumed. Then, we used the intrinsic conditional auto-regressive (CAR) specification with improper prior, the CAR model with proper prior, and also the Besag-York-Mollie (BYM) model \cite{Besag1991}. 

Specifically, the CAR specification for the spatial effect for a specific year is defined by:
$$\theta_{i,(year)}|\theta_{j,(year),\tau_\theta} \sim N\left( \frac{1}{n_i} \sum_{j \sim i} \theta_{j,(year)}, \frac{1}{\tau_\theta n_i}  \right),$$
where $\tau_\theta$ is the conditional precision, $j\sim i$ denotes that $\boldsymbol{W}_{ij} = 1$, and $n_i$ is the numbers of neighbors, according to the definition of the two types of proximity matrix. The proper CAR model is obtained by adding a positive quantity $d$ to $n_i$, whereas the BYM model is obtained by adding an unstructured random effect per municipality. For more details, see \cite{Bivand2015}. The R packages \verb|dlnm|~\cite{Gasparrini2011} and \verb|INLA| \cite{Rue2009} were used for all calculations. 

Finally, we performed two simple forecasting procedures to compare them with the proposed model in terms of predictive skill. Both methods use the training set for the last five years (2015-2019). First, naïve forecasting is performed using a simple monthly mean $\hat{RR_{it}}$ for the last five years. This method is easy to compute, but the prediction interval can only be calculated by assuming strong assumptions, such as independence; thus, only $NMRSE$ can be computed. 

Second, as an alternative, we estimated the model \eqref{eq:model} with
\begin{equation*}
	\log RR_{it}=\alpha, 
\end{equation*}
as a Negative Binomial null model so that the prediction uncertainty can be computed in this case. The main objective of performing these two prediction procedures is to compare how the proposed model outperforms these two simple algorithms by using a more complex structure composed of the covariates and spatial random effect.

\subsection{Model selection and prediction}

Previous studies \cite{garcia2023wavelet,Barboza2023} have shown that these climatic covariates are essential to predict dengue incidence. To begin the calibration, a training period is chosen to fit the model \ref{eq:model} using different combinations of covariates and spatio-temporal configurations of the model. First, the DLNM framework allows us to choose different combinations of maximum and minimum lags of historical information on the covariates. Moreover, the basis was chosen to be nonlinear and linear for all covariates. Finally, four spatial structures were fitted: independent, CAR, proper CAR, and BYM models.

The best model was chosen by comparing all fitted models by different criteria. The deviance information criteria (DIC) and the mean cross-validation (CV) log score are calculated for each model. The DIC is a measure that contemplates the model's precision and complexity. At the same time, the CV log score is a criterion that measures the model's predictive capacity, letting one data out at a time.

Finally, two metrics to compare the predictive performance of each model are computed. The normalized Mean-Squared Error ($NRMSE$):
\begin{align*}
	NRMSE = \sqrt{\frac{1}{m\overline{ RR}}\sum_{t=1}^m(RR_t-\widehat {RR}_t)^2},
\end{align*}
where $m$ is the number of months in the testing period, $\overline{RR}$ is the mean relative risk over the same period, and $\widehat{RR}$ is the estimated relative risk according to any of the two models. The normalized Interval Score at $\alpha$ level ($NIS_{\alpha}$) is the normalized version of the Interval Score (see~\cite{GoodProbabilityAssessors} and~\cite{Gneiting2007a}):
\begin{align*}
	NIS_{\alpha}=\frac{1}{m\overline{RR}} \sum_{t=1}^m\left[(U_t-L_t)+\frac{2}{1-\alpha}(L_t-RR_t)\cdot 1_{RR_t<L_t}+\frac{2}{1-\alpha}(RR_t-U_t)\cdot 1_{RR_t>U_t}\right],
\end{align*}
where $U_t$ and $L_t$ are the upper and lower limits of the prediction interval, respectively, the latter metric is more complete in evaluating the models' predictive capacity when the uncertainty is summarized through a predictive interval~\cite{Gneiting2007a}. It has been used in previous predictive studies on dengue fever in Costa Rica (see \cite{vazquez2020,Barboza2023}). We use the normalized version of RMSE and IS because we can compare different locations regardless of the scale of their relative risk.

\section{Results}

We fitted the model \eqref{eq:model} specifying different structures to the dengue data using all climate covariates. The training period consists of monthly data from January 2000 to December 2020, and the testing period from January 2021 to March 2021.

Regarding the DLNM framework, specifying the maximum and minimum lags to incorporate historical exposure of climate covariates is challenging. We set the maximum and minimum lags as 12 and 3 lags, respectively, so that we can predict the dengue data for up to three months. We also fitted the models using maximum and minimum lags of 12 and 0, respectively, and the difference in predictive precision is insignificant with respect to the first alternative. Table \ref{tab:fitting} presents all models' DIC and CV log-scores with the smallest values per metric in bold. 
\begin{table}[htp!]
\caption{\label{tab:fitting} Comparison of the models according to Deviance information criterion (DIC) and mean cross-validation (CV) log-score.}
\centering
\begin{threeparttable}
\begin{tabular}[t]{llccc}
\toprule
DLNM & Proximity matrix & Spatial structure & DIC & CV log-score\\
\midrule
\multirow{7}{*}{Linear\tnote{*}} &  & Independent & 57135.37 & 3.8710\\
 & \multirow{3}{*}{Neighbor} & CAR & 54256.47 & 3.6872\\
 &  & proper CAR & \textbf{52628.40} & \textbf{3.5774}\\
 &  & BYM & 52632.24 & 3.5784\\
 & \multirow{3}{*}{Distance} & CAR & 53416.92 & 3.6264\\
 &  & proper CAR & 52633.29 & 3.5787\\
 &  & BYM & 52636.92 & 3.5787\\
\midrule
\multirow{7}{*}{Non-linear\tnote{*}} &  & Independent & 53429.63 & 3.8756\\
 & \multirow{3}{*}{Neighbor} & CAR & 50640.66 & 3.6838\\
 &  & proper CAR & \textbf{49438.81} & \textbf{3.5954}\\
 &  & BYM & 49461.01 & 3.5977\\
 & \multirow{3}{*}{Distance} & CAR & 54674.34 & 3.9653\\
 &  & proper CAR & 49468.89 & 3.5985\\
 &  & BYM & 49456.27 & 3.5971\\
\bottomrule
\end{tabular}
  \begin{tablenotes}
    \item[*] The best model for each DLNM specification is marked in bold.
  \end{tablenotes}
\end{threeparttable}
\end{table}
First, it is clear that, in terms of the goodness of fit, the models with the spatial structure are better than those assuming independent spatial structures. Then, we can see that the differences in DIC and CV log scores are minimal when we compare the linear and non-linear (B-splines) DLNM framework. 

Finally, to guarantee an acceptable balance between the complexity of the models and the predictive precision over all the locations using the DIC and CV log-scores, we chose as the best-fitted model the proper CAR model with linear DLNM and with neighbor proximity matrix.

Table \ref{tab:predictive_metric} summarizes the predictive metrics ($NRMSE$ and $NIS_{0.05}$) of the training and testing periods per municipality for the best model compared to the baseline model with an independent spatial structure. 
\begin{table}[htp!]
\small
\centering
	\caption{\label{tab:predictive_metric} Predictive metrics of training and testing data set of the selected model}
\begin{threeparttable}
\begin{tabular}[t]{lrrrrrrrr}
\toprule
\multicolumn{1}{c}{ } & \multicolumn{4}{c}{Training set} & \multicolumn{4}{c}{Testing set} \\
\cline{2-5} \cline{6-9}
\multicolumn{1}{c}{Municipality} & \multicolumn{2}{c}{Independent} & \multicolumn{2}{c}{Proper CAR} & \multicolumn{2}{c}{Independent} & \multicolumn{2}{c}{Proper CAR} \\
\cline{2-3} \cline{4-5} \cline{6-7} \cline{8-9}
 & $NRMSE$ & $NIS_{95}$ & $NRMSE$ & $NIS_{0.05}$ & $NRMSE$ & $NIS_{0.05}$ & $NRMSE$ & $NIS_{0.05}$\\
\midrule
Alajuela & 0.7815 & 14.5297 & 0.3982 & 5.3710 & 0.0750 & 1.0905 & 0.0416 & 1.6417\\
Alajuelita & 0.3090 & 23.2238 & 0.2175 & 13.1177 & 0.4515 & 22.8135 & 0.0515 & 2.0922\\
Atenas & 4.6100 & 24.7736 & 2.5706 & 10.1453 & 5.1453 & 87.6173 & 0.3283 & 10.2199\\
Cañas & 10.1008 & 24.3243 & 5.7069 & 12.3277 & 5.8803 & 60.7800 & 0.4829 & 6.8129\\
Carrillo & 4.3786 & 15.9598 & 3.5404 & 9.1421 & 0.9860 & 13.2041 & 0.4175 & 2.5945\\
Corredores & 5.0620 & 25.1830 & 2.5155 & 8.6126 & 1.6824 & 17.0115 & 0.5697 & 1.9358\\
Desamparados & 0.2351 & 18.6300 & 0.1466 & 8.2922 & 0.0282 & 3.0978 & 0.0142 & 3.0085\\
Esparza & 4.7287 & 17.4775 & 2.6296 & 8.4081 & 1.1899 & 16.1291 & 0.1849 & 2.1748\\
Garabito & 38.4940 & 29.0191 & 5.2545 & 9.8071 & 28.8475 & 232.7807 & 0.4258 & 12.9528\\
Golfito & 3.7245 & 23.5563 & 1.7734 & 8.2174 & 2.0078 & 35.7736 & 0.3499 & 8.5601\\
Guacimo & 1.9057 & 13.8120 & 1.0844 & 4.4132 & 3.0709 & 18.1805 & 2.0633 & 4.8628\\
La Cruz & 6.2484 & 26.6755 & 4.2779 & 14.3090 & 6.0351 & 118.6721 & 0.5029 & 14.7439\\
Liberia & 3.6829 & 23.1675 & 2.0260 & 10.2626 & 4.0913 & 92.4202 & 0.5693 & 16.2507\\
Limon & 2.9640 & 17.1259 & 1.6025 & 6.7593 & 1.3724 & 11.2322 & 0.1034 & 1.7042\\
Matina & 5.1143 & 19.9086 & 2.7162 & 7.4890 & 58.0750 & 192.2903 & 0.1232 & 3.0869\\
Montes de Oro & 5.9220 & 21.8260 & 4.1289 & 12.2332 & 11.4909 & 126.5836 & 1.2106 & 19.5139\\
Nicoya & 2.8036 & 20.5059 & 2.0672 & 10.1305 & 3.6262 & 101.6524 & 0.1894 & 10.2497\\
Orotina & 13.7161 & 17.5376 & 4.0681 & 6.0472 & 1.0599 & 6.7613 & 0.4888 & 1.5221\\
Osa & 5.9295 & 33.7053 & 4.2208 & 17.2559 & 1.1235 & 13.0806 & 0.5758 & 1.7891\\
Parrita & $1.24 \times 10^9$ & 24483.9181 & 13.7622 & 9.8340 & 4.4446 & 46.3670 & 0.6773 & 7.9705\\
Perez Zeledón & 2.0103 & 30.0358 & 0.8733 & 9.4268 & 1.2543 & 20.3380 & 0.1783 & 1.3942\\
Pococí & 2.2524 & 12.7658 & 1.3001 & 6.1328 & 1.0524 & 12.0212 & 0.1284 & 1.8534\\
Puntarenas & 1.5802 & 12.8416 & 0.9303 & 4.6933 & 1.3263 & 23.9665 & 0.1646 & 1.7646\\
Quepos & 56.8344 & 37.6499 & 10.0908 & 12.9868 & 36.9331 & 257.7509 & 1.1780 & 23.2153\\
San Jose & 0.2105 & 12.6814 & 0.1328 & 5.2650 & 0.0215 & 1.2409 & 0.0155 & 2.3248\\
Santa Ana & 0.7837 & 21.9311 & 0.5860 & 12.6213 & 0.6759 & 43.5264 & 0.3576 & 15.4854\\
SantaCruz & 35.1198 & 37.3329 & 8.4762 & 12.7070 & 6.5905 & 91.5896 & 0.3027 & 3.0635\\
Sarapiquí & 7.9218 & 22.2392 & 2.8096 & 6.9807 & 0.3880 & 4.9820 & 0.1047 & 1.9942\\
Siquirres & 2.1184 & 13.9371 & 1.4077 & 6.6140 & 2.7214 & 19.0646 & 0.3564 & 1.7977\\
Talamanca & 4.9193 & 21.1751 & 2.2374 & 8.0522 & 0.1113 & 0.7642 & 0.7989 & 1.3152\\
Turrialba & 2.5905 & 25.2386 & 1.9572 & 15.8268 & 1.0122 & 20.0694 & 0.2614 & 1.8489\\
Upala\tnote{1} & 1.2744 & 21.7159 & 0.9203 & 12.2963 & - & - & - & -\\
\bottomrule
\end{tabular}
  \begin{tablenotes}
    \item[1] $NRMSE$ and $NIS_{0.05}$ of the testing set for Upala are not shown since the observed relative risks are zero.
  \end{tablenotes}
\end{threeparttable}
\end{table}

We can observe how spatial information can contribute to obtaining more precise predictions by comparing those two modeling alternatives. The most noticeable municipality Parrita has a particular behavior. The baseline model with an independent spatial structure cannot predict training and testing periods. In contrast, our selected model with proper CAR spatial structure can substantially reduce its prediction metrics. Moreover, the precision of the proposed model performs better than the simple forecasting procedures (see Table A.1. in Supplementary Material).

Close to Parrita, two municipalities with moderately high predictive metrics are Garabito and Quepos, located on the country's central pacific coast. We suspect that the difficulty of predicting these particular areas is mainly because they are touristic, and dengue cases in these areas are likely underreported.

\begin{figure}[!htp]
	\centering
	\includegraphics[scale=0.3]{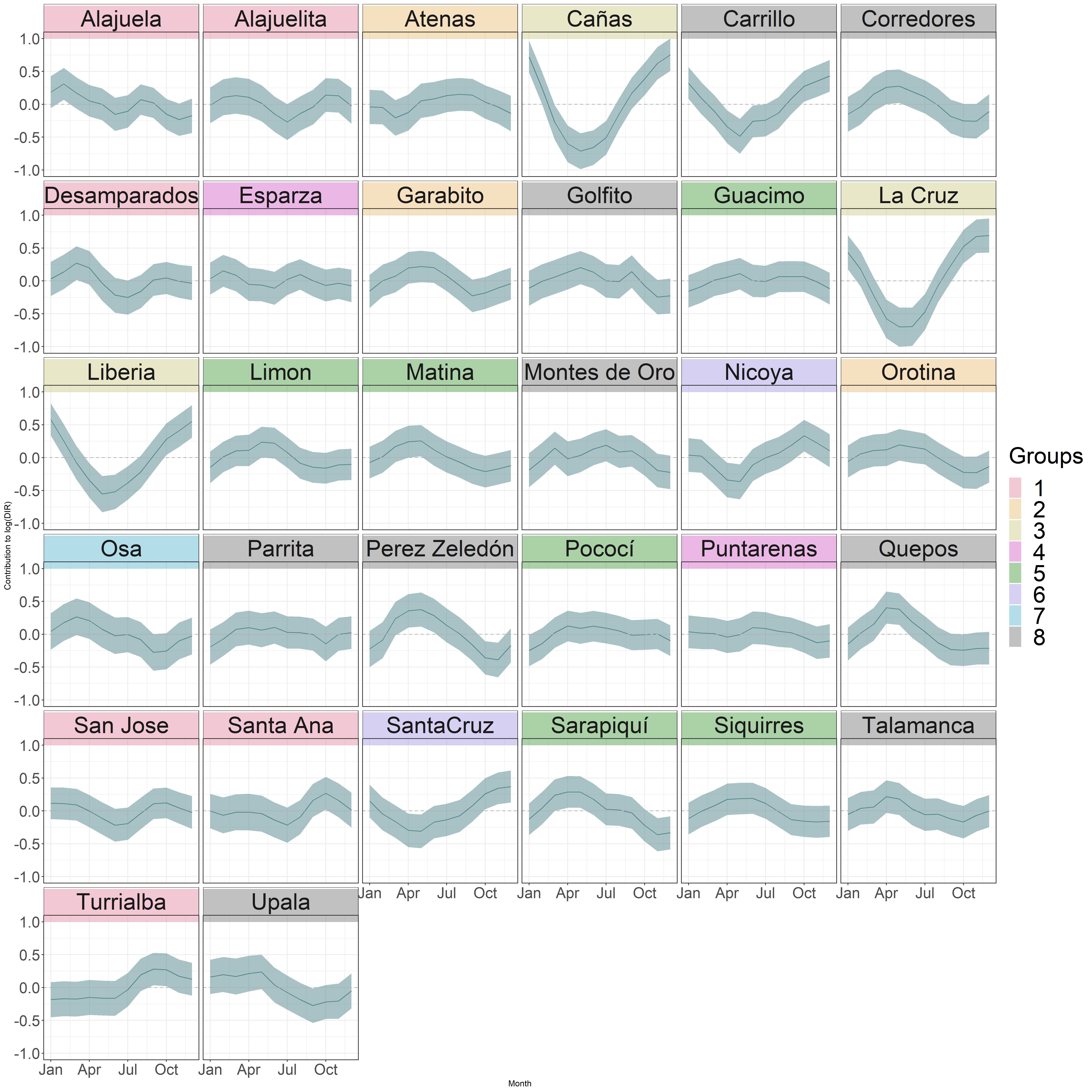}
	\caption{Posterior mean and 95\% credible interval of municipality-specific monthly random effects.}
	\label{fig:temporal.random}
\end{figure}

\begin{figure}[!htp]
	\centering
	\includegraphics[scale=0.2]{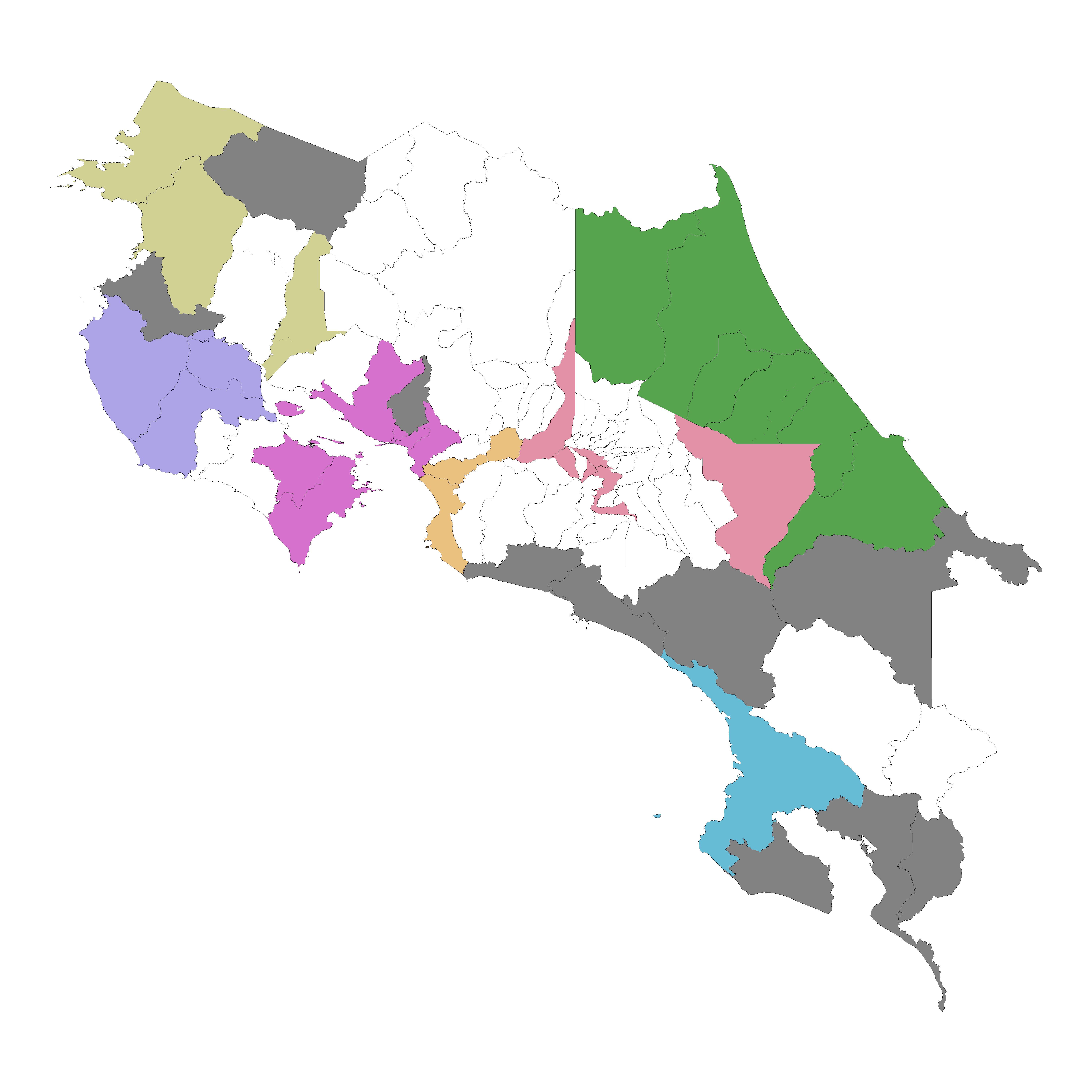}
	\caption{Illustration of eight groups with similar temporal behavior.}
	\label{fig:temporal.random.map}
\end{figure}

Besides the climatic covariate's contribution to the model, temporal and spatial random effects ($\phi_{i,(month)}$ and $\theta_{i,(year)}$) are also important factors in modeling dengue behavior because they provide temporal or spatial information which is not observable through the selected covariates. Figure \ref{fig:temporal.random} shows the behavior of municipality-specific monthly random effects. Municipalities with similar temporal behavior are identified as Groups 1 to 7, and the last group consists of municipalities that do not have a specific behavior. Later on, the geographic location of these groups of municipalities is illustrated in Figure \ref{fig:temporal.random.map}. We observe that the information provided by the temporal random effects mostly agrees with the temporal variation due to the microclimates in the country.

\begin{figure}[!htp]
	\centering
	\includegraphics[scale=0.3]{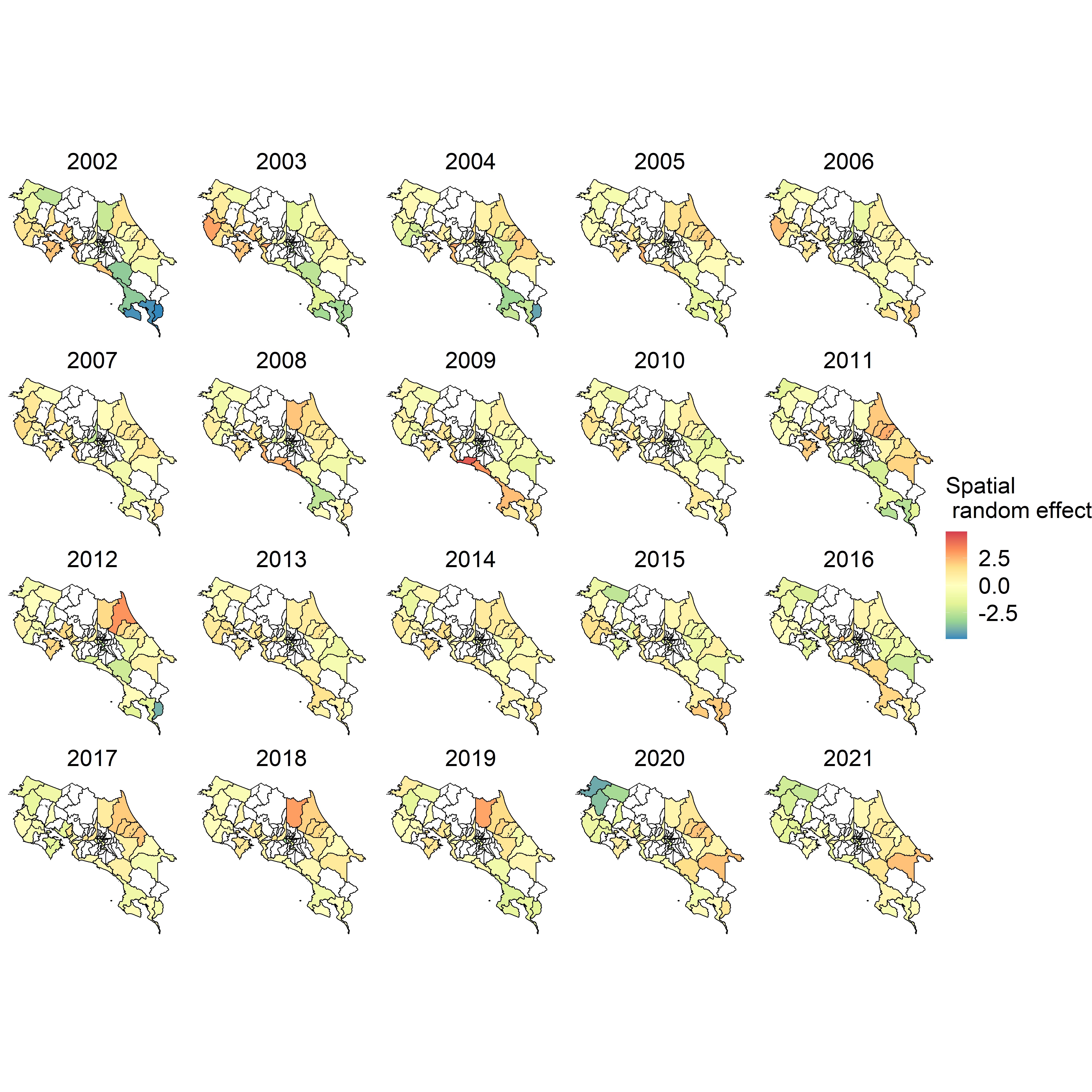}
	\caption{Contribution of year-specific spatial random effect to dengue log relative risk.}
	\label{fig:spatial.random}
\end{figure}		

\newpage
On the other hand, the spatial random effect for each year is illustrated in Figure \ref{fig:spatial.random}. We observe that there are clusters of cantons for different years that present a higher or lower log of relative risk. For instance, the south pacific part of the country in 2002 and the north pacific region presented a lower contribution of log of relative risk, while the central pacific in 2009 presented a high contribution of log of relative risk.

It is important to notice that this spatial random effect is the variation in log of relative risk that is not captured by the climatic covariates. Still, somehow they can be modeled by the neighbor's spatial structure. It is known that dengue is a complex phenomenon that involves not only climatic factors but also socioeconomic components and human mobility. Furthermore, there are also outbreaks in certain municipalities during different periods that are not captured by the climatic covariates in this study. In this way, this model is more complete compared to the baseline model assuming independent spatial structure.

\begin{figure}[!htp]
	\centering
	\includegraphics[scale=0.1]{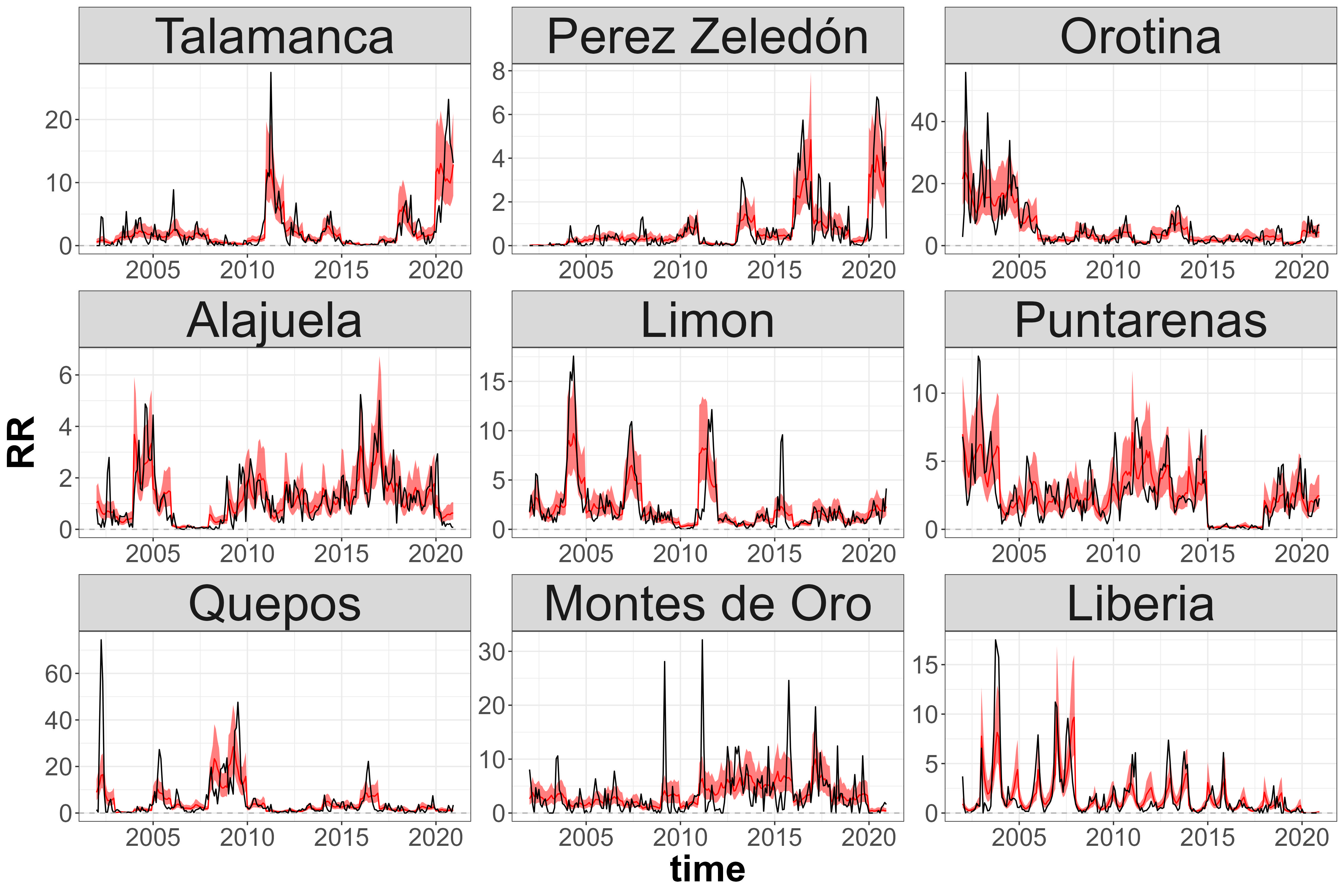}
	\caption{Observed (black) and 95\% posterior predictive dengue relative risks (red) over the training period. Upper six panels: best municipalities according to NIS metric. Lower three panels: worst municipalities according to NIS metric.}
	\label{fig:RR.prediction.in}
\end{figure}

\begin{figure}[!htp]
	\centering
	\includegraphics[scale=0.1]{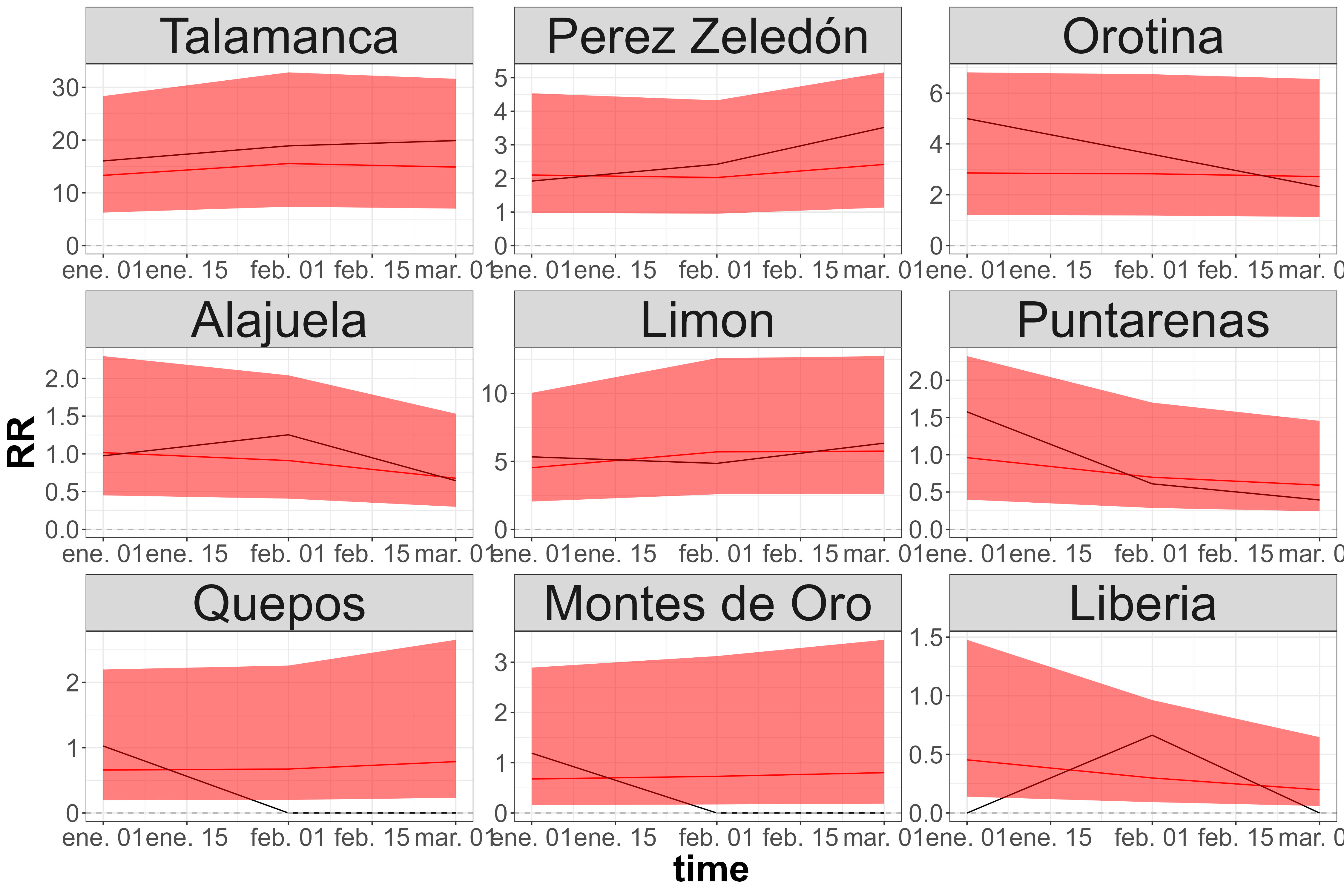}
	\caption{Observed (black) and 95\% posterior predictive dengue relative risks (red) over the testing period. Upper six panels: best municipalities according to NIS metric. Lower three panels: worst municipalities according to NIS metric.}
	\label{fig:RR.prediction.out}
\end{figure}

The dengue prediction can be visualized by using temporal or spatial dimensions. For the temporal dimension, Figure \ref{fig:RR.prediction.in} shows the 95\% posterior predictive dengue relative risk in the training period of the best six municipalities and the three worst municipalities according to $NIS_{0.05}$ during the testing period. In addition, Figure \ref{fig:RR.prediction.out} presents the \% posterior predictive dengue relative risk of the same municipalities during the testing period. In general, the behaviors of dengue relative risks can be precisely modeled, and the prediction distribution can also capture the observed RR during the testing periods. It should be noted that the predictive uncertainty is positively asymmetric due to the asymmetric nature of relative risks and the model contemplates the log of relative risks.

\begin{figure}[!htp]
\begin{subfigure}[t]{0.5\linewidth}
	\includegraphics[scale=0.05]{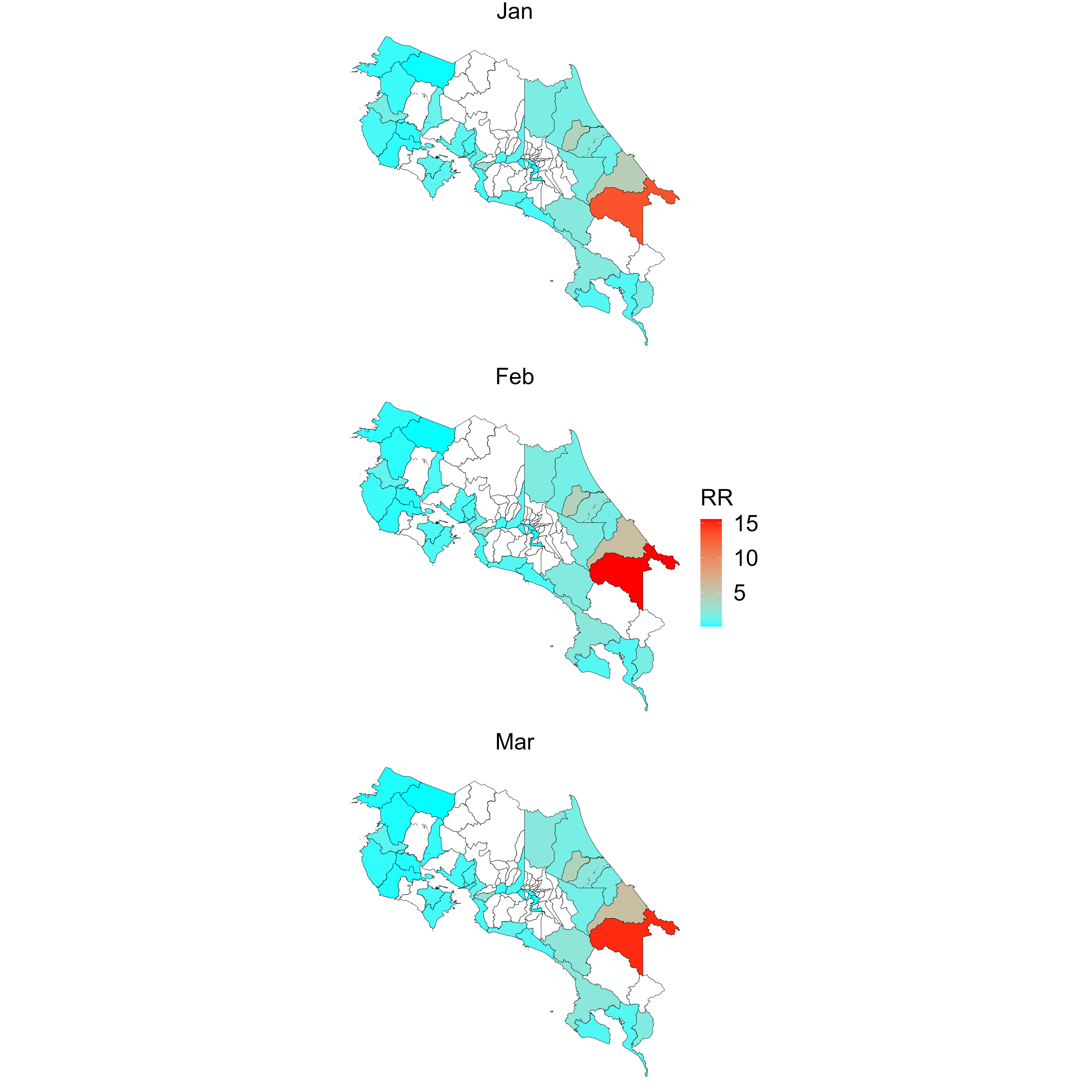}
	\caption{Relative risk prediction.}
	\label{fig:prediction.map.out}
\end{subfigure}\quad
\begin{subfigure}[t]{0.5\linewidth}
	\includegraphics[scale=0.05]{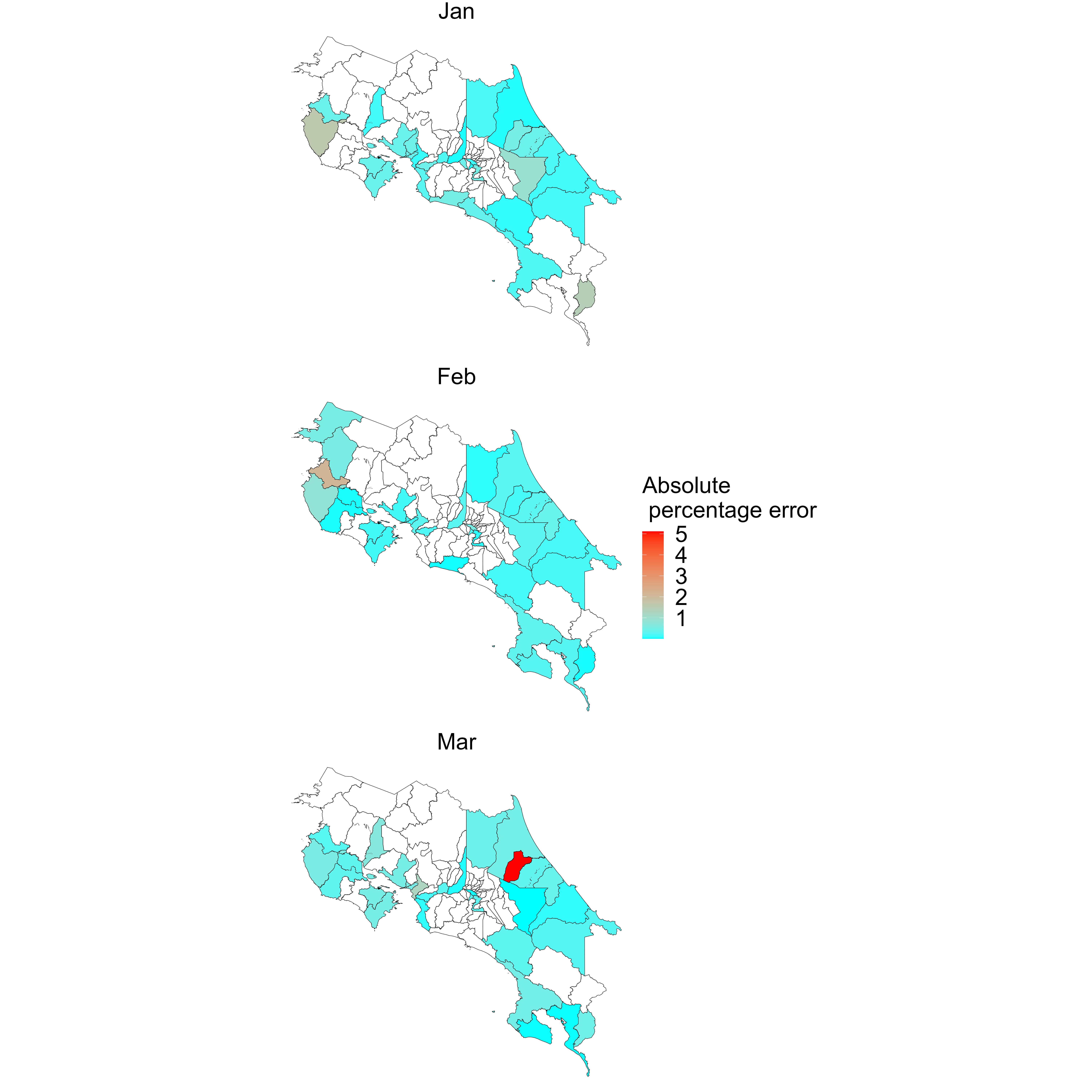}
	\caption{Absolute percentage error.}
	\label{fig:abe.map.out}
\end{subfigure}
\hfill
\caption{Relative risk prediction and their absolute percentage error from January to March 2020 for 81 municipalities in Costa Rica.}
\label{fig:prediction.out.map}
\end{figure}

For the spatial dimension, Figure \ref{fig:prediction.out.map} presents the posterior predictive dengue relative risk mean and their absolute percentage error for each municipality and month (January, February, and March). These maps allow us to visualize regions with higher dengue relative risks in a specific month. Furthermore, Figure \ref{fig:abe.map.out} shows that the absolute percentage error is uniform across the region for the testing period.

For the training set, similar behavior is visualized. Maps of the posterior mean of relative risks for three selected years (2002, 2011, 2020) and maps of absolute percentage error for the same years are shown in Sections C and D of the Supplementary Material.

\section{Discussion}
Epidemiological models provide a crucial tool to help public health authorities and policymakers to allocate limited resources effectively and efficiently. Using these predictive models helps understand disease dynamics, the potential impact on the population, and the health system's response to an outbreak. These models can be improved by including demographics, genetics, and environmental variables to make more accurate predictions.

One of the challenges in developing and using epidemiological models is obtaining high-quality data. A robust and comprehensive data collection system is essential to provide accurate input for the models. In addition, the data must be integrated and processed consistently and timely to ensure that the models are accurate and up-to-date. Data sources include health records, laboratory results, surveillance systems, and household surveys. The output of these models can help in decision-making processes concerning control purposes and surveillance methods and hopefully also as good predictive tools. Prediction forms part of surveillance systems, and more specifically in early warning systems~\cite{racloz2012surveillance}. 


The effect of climate variability and climate change on dengue transmission is complex, nonlinear, and often delayed by several weeks to months~\cite{naish2014climate,wen2006spatial}. Bringing together spatio-temporal patterns of dengue transmission compatible with long-term data on climate and other socio-ecological changes in mathematical and statistical models could improve projections of dengue risks and disease control.

Based on the results of fitting the model \eqref{eq:model} to the dengue data, it is clear that incorporating all climate covariates greatly improved the model's performance in accurately predicting the number of dengue cases. Using different structures in the model allowed for a deeper understanding of the relationships between dengue transmission and climate factors. The model's results demonstrate its potential to be a useful tool for decision-making processes in disease control and surveillance methods. Furthermore, the successful application of the model to monthly data from January 2000 to March 2021 highlights its potential for future predictions and early warning systems.

The spatio-temporal random components in the model make us aware that more structural information, such as human mobility and socio-economic factors, could be considered to obtain better predictive results. Access to these data is challenging, and we are working on it in future investigations.

Finally, by providing accurate predictions, decision-makers can respond more effectively to outbreaks and implement strategies to reduce the impact of the disease on the population. Thus, helping health authorities optimize the typically scarce resources. To maximize their effectiveness, models must be based on high-quality data, continuously updated, and validated jointly with public health officials to reflect environmental changes and other factors, including social determinants, that may impact disease transmission. The continued development of these models will play a critical role in the fight against dengue and other infectious diseases.

\bibliographystyle{unsrt}
\bibliography{references}

\begin{thebibliography}{10}

\bibitem{WHO:online}
World Health~Organization (WHO).
\newblock Dengue and sever dengue, August 2022.

\bibitem{medlock2006analysis}
Jolyon~M Medlock, David Avenell, Iain Barrass, and Steve Leach.
\newblock Analysis of the potential for survival and seasonal activity of aedes
  albopictus (diptera: Culicidae) in the united kingdom.
\newblock {\em Journal of Vector Ecology}, 31(2):292--304, 2006.

\bibitem{romi2006cold}
Roberto Romi, Francesco Severini, and Luciano Toma.
\newblock Cold acclimation and overwintering of female aedes albopictus in
  roma.
\newblock {\em Journal of the American Mosquito Control Association},
  22(1):149--151, 2006.

\bibitem{outammassine2022global}
Abdelkrim Outammassine, Said Zouhair, and Souad Loqman.
\newblock Global potential distribution of three underappreciated arboviruses
  vectors (aedes japonicus, aedes vexans and aedes vittatus) under current and
  future climate conditions.
\newblock {\em Transboundary and Emerging Diseases}, 69(4):e1160--e1171, 2022.

\bibitem{murray2013epidemiology}
Natasha Evelyn~Anne Murray, Mikkel~B Quam, and Annelies Wilder-Smith.
\newblock Epidemiology of dengue: past, present and future prospects.
\newblock {\em Clinical epidemiology}, 5:299, 2013.

\bibitem{massad2018estimating}
Eduardo Massad, Marcos Amaku, Francisco Antonio~Bezerra Coutinho,
  Claudio~Jos{\'e} Struchiner, Marcelo~Nascimento Burattini, Kamran Khan, Jing
  Liu-Helmersson, Joacim Rockl{\"o}v, Moritz~UG Kraemer, and Annelies
  Wilder-Smith.
\newblock Estimating the probability of dengue virus introduction and secondary
  autochthonous cases in europe.
\newblock {\em Scientific reports}, 8(1):1--12, 2018.

\bibitem{lopez2016modeling}
Luis~Fernandez Lopez, Marcos Amaku, Francisco Antonio~Bezerra Coutinho, Mikkel
  Quam, Marcelo~Nascimento Burattini, Claudio~Jos{\'e} Struchiner, Annelies
  Wilder-Smith, and Eduardo Massad.
\newblock Modeling importations and exportations of infectious diseases via
  travelers.
\newblock {\em Bulletin of mathematical biology}, 78(2):185--209, 2016.

\bibitem{gubler2012economic}
Duane~J Gubler.
\newblock The economic burden of dengue.
\newblock {\em The American journal of tropical medicine and hygiene},
  86(5):743, 2012.

\bibitem{wang2022global}
Hao Wang, Shaohua Zhao, Shengjun Wang, Yue Zheng, Shaohua Wang, Hui Chen,
  Jiaojiao Pang, Juan Ma, Xiaorong Yang, and Yuguo Chen.
\newblock Global magnitude of encephalitis burden and its evolving pattern over
  the past 30 years.
\newblock {\em Journal of Infection}, 84(6):777--787, 2022.

\bibitem{messina2019current}
Jane~P Messina, Oliver~J Brady, Nick Golding, Moritz~UG Kraemer, GR~Wint,
  Sarah~E Ray, David~M Pigott, Freya~M Shearer, Kimberly Johnson, Lucas Earl,
  et~al.
\newblock The current and future global distribution and population at risk of
  dengue.
\newblock {\em Nature microbiology}, 4(9):1508--1515, 2019.

\bibitem{yang2021global}
Xiaorong Yang, Mikkel~BM Quam, Tongchao Zhang, and Shaowei Sang.
\newblock Global burden for dengue and the evolving pattern in the past 30
  years.
\newblock {\em Journal of travel medicine}, 28(8):taab146, 2021.

\bibitem{morin2013climate}
Cory~W Morin, Andrew~C Comrie, and Kacey Ernst.
\newblock Climate and dengue transmission: evidence and implications.
\newblock {\em Environmental health perspectives}, 121(11-12):1264--1272, 2013.

\bibitem{naish2014climate}
Suchithra Naish, Pat Dale, John~S Mackenzie, John McBride, Kerrie Mengersen,
  and Shilu Tong.
\newblock Climate change and dengue: a critical and systematic review of
  quantitative modelling approaches.
\newblock {\em BMC infectious diseases}, 14(1):1--14, 2014.

\bibitem{campbell2015weather}
Karen~M Campbell, Kristin Haldeman, Chris Lehnig, Cesar~V Munayco, Eric~S
  Halsey, V~Alberto Laguna-Torres, Mart{\'\i}n Yagui, Amy~C Morrison, Chii-Dean
  Lin, and Thomas~W Scott.
\newblock Weather regulates location, timing, and intensity of dengue virus
  transmission between humans and mosquitoes.
\newblock {\em PLoS neglected tropical diseases}, 9(7):e0003957, 2015.

\bibitem{watts1986effect}
Douglas~M Watts, Donald~S Burke, Bruce~A Harrison, Richard~E Whitmire, and
  Ananda Nisalak.
\newblock Effect of temperature on the vector efficiency of aedes aegypti for
  dengue 2 virus.
\newblock Technical report, ARMY MEDICAL RESEARCH INST OF INFECTIOUS DISEASES
  FORT DETRICK MD, 1986.

\bibitem{tun2000effects}
W~Tun-Lin, TR~Burkot, and BH~Kay.
\newblock Effects of temperature and larval diet on development rates and
  survival of the dengue vector aedes aegypti in north queensland, australia.
\newblock {\em Medical and veterinary entomology}, 14(1):31--37, 2000.

\bibitem{rueda1990temperature}
LM~Rueda, KJ~Patel, RC~Axtell, and RE~Stinner.
\newblock Temperature-dependent development and survival rates of culex
  quinquefasciatus and aedes aegypti (diptera: Culicidae).
\newblock {\em Journal of medical entomology}, 27(5):892--898, 1990.

\bibitem{sarfraz2012analyzing}
Muhammad~Shahzad Sarfraz, Nitin~K Tripathi, Taravudh Tipdecho, Thawisak
  Thongbu, Pornsuk Kerdthong, and Marc Souris.
\newblock Analyzing the spatio-temporal relationship between dengue vector
  larval density and land-use using factor analysis and spatial ring mapping.
\newblock {\em BMC public health}, 12(1):1--19, 2012.

\bibitem{dieng2012household}
Hamady Dieng, Abu~Hassan Ahmad, Jazem~A Mahyoub, Abdulhafis~M Turkistani,
  Hatabbi Mesed, Salah Koshike, Tomomitsu Satho, MR~Che Salmah, Hamdan Ahmad,
  Wan~Fatma Zuharah, et~al.
\newblock Household survey of container--breeding mosquitoes and climatic
  factors influencing the prevalence of aedes aegypti (diptera: Culicidae) in
  makkah city, saudi arabia.
\newblock {\em Asian Pacific journal of tropical biomedicine}, 2(11):849--857,
  2012.

\bibitem{van2005spatial}
Birgit~HB Van~Benthem, Sophie~O Vanwambeke, Nardlada Khantikul, Chantal
  Burghoorn-Maas, Kamolwan Panart, Linda Oskam, Eric~F Lambin, and Pradya
  Somboon.
\newblock Spatial patterns of and risk factors for seropositivity for dengue
  infection.
\newblock {\em The American journal of tropical medicine and hygiene},
  72(2):201--208, 2005.

\bibitem{barrera2006ecological}
Roberto Barrera, Manuel Amador, and Gary~G Clark.
\newblock Ecological factors influencing aedes aegypti (diptera: Culicidae)
  productivity in artificial containers in \uppercase{S}alinas,
  \uppercase{P}uerto \uppercase{R}ico.
\newblock {\em Journal of medical entomology}, 43(3):484--492, 2006.

\bibitem{mateus2011predictors}
Julio~C{\'e}sar Mateus and Gabriel Carrasquilla.
\newblock Predictors of local malaria outbreaks: an approach to the development
  of an early warning system in colombia.
\newblock {\em Mem{\'o}rias do Instituto Oswaldo Cruz}, 106:107--113, 2011.

\bibitem{eidson2001dead}
Millicent Eidson, Laura Kramer, Ward Stone, Yoichiro Hagiwara, Kate Schmit, New
  York State West Nile Virus Avian~Surveillance Team, et~al.
\newblock Dead bird surveillance as an early warning system for west nile
  virus.
\newblock {\em Emerging infectious diseases}, 7(4):631, 2001.

\bibitem{VBORNET:online}
European~Centre for Disease~Prevention and Control.
\newblock Vbornet--european network for arthropod vector surveillance for human
  public health, August 2022.

\bibitem{SitioWeb58:online}
Ministerio de~Salud.
\newblock Sitio web del ministerio de salud de costa rica. bienvenido, 2022.

\bibitem{vazquez2020}
Paola V\'asquez, Antonio Lor\'ia, Fabio Sanchez, and Luis~A. Barboza.
\newblock {Climate-driven Statistical Models as effective predictors of local
  Dengue incidence in Costa Rica: a Generalized Additive Model and Random
  Forest Approach}.
\newblock {\em Revista de Matematica: Teoria y Aplicaciones}, 27(1):1--21,
  2020.

\bibitem{garcia2023wavelet}
Yury~E. Garcia, Shu-Wei Chou-Chen, Luis~A. Barboza, Maria~L. Daza-Torres,
  J.~Cricelio Montesinos-Lopez, Paola Vasquez, Juan~G. Calvo, Miriam Nuno, and
  Fabio Sanchez.
\newblock Common patterns between dengue cases, climate, and local
  environmental variables in costa rica: A wavelet approach, 2023.

\bibitem{minsa:online}
Ministerio de~Salud.
\newblock Sitio web del ministerio de salud de costa rica. bienvenido, 2022.

\bibitem{FunkChris2015Tchi}
Chris Funk, Pete Peterson, Martin Landsfeld, Diego Pedreros, James Verdin,
  Shraddhanand Shukla, Gregory Husak, James Rowland, Laura Harrison, Andrew
  Hoell, and Joel Michaelsen.
\newblock The climate hazards infrared precipitation with stations--a new
  environmental record for monitoring extremes.
\newblock {\em Scientific data}, 2(1):150066--150066, 2015.

\bibitem{CPC-SSTA}
NOAA.
\newblock Climate prediction center.
\newblock
  \url{https://www.cpc.ncep.noaa.gov/data/indices/ersst5.nino.mth.91-20.ascii}.
\newblock Accessed: 2022-05-01.

\bibitem{TuckPhillips}
Sean~L. Tuck, Helen~R.P. Phillips, Rogier~E. Hintzen, J{"o}rn~P.W. Scharlemann,
  Andy Purvis, and Lawrence~N. Hudson.
\newblock Modistools -- downloading and processing modis remotely sensed data
  in r.
\newblock {\em Ecology and Evolution}, 4(24):4658--4668, 2014.

\bibitem{EnfieldDavidB1999Huit}
David~B Enfield, Alberto~M Mestas-Nuñez, Dennis~A Mayer, and Luis Cid-Serrano.
\newblock How ubiquitous is the dipole relationship in tropical atlantic sea
  surface temperatures?
\newblock {\em Journal of Geophysical Research: Oceans}, 104(C4):7841--7848,
  1999.

\bibitem{Hidalgo2017-hr}
H~G Hidalgo, E~J Alfaro, and B~Quesada-Montano.
\newblock Observed (1970--1999) climate variability in central america using a
  high-resolution meteorological dataset with implication to climate change
  studies.
\newblock {\em Clim. Change}, 141(1):13--28, 2017.

\bibitem{Gasparrini2010}
A.~Gasparrini, B.~Armstrong, and M.~G. Kenward.
\newblock Distributed lag non-linear models.
\newblock {\em Statistics in Medicine}, 29(21):2224--2234, 2010.

\bibitem{Gasparrini2014}
Antonio Gasparrini.
\newblock Modeling exposure–lag–response associations with distributed lag
  non-linear models.
\newblock {\em Statistics in Medicine}, 33(5):881--899, 2014.

\bibitem{Besag1991}
Julian Besag, Jeremy York, and Annie Mollié.
\newblock Bayesian image restoration, with two applications in spatial
  statistics.
\newblock {\em Annals of the Institute of Statistical Mathematics},
  43(1):1--20.

\bibitem{Bivand2015}
Roger Bivand, Virgilio Gómez-Rubio, and Håvard Rue.
\newblock Spatial data analysis with r-inla with some extensions.
\newblock {\em Journal of Statistical Software}, 63(20):1–31, 2015.

\bibitem{Gasparrini2011}
A.~Gasparrini.
\newblock Distributed lag linear and non-linear models in {R}: the package
  {dlnm}.
\newblock {\em Journal of Statistical Software}, 43(8):1--20, 2011.

\bibitem{Rue2009}
H{\^{a}}vard Rue, Sara Martino, and Nicolas Chopin.
\newblock {Approximate Bayesian Inference for Latent Gaussian Models by Using
  Integrated Nested Laplace Approximations}.
\newblock {\em Journal of the Royal Statistical Society . Series B (
  Methodological )}, 71(2):319--392, 2009.

\bibitem{Barboza2023}
Luis~A. Barboza, Shu-Wei Chou-Chen, Paola Vásquez, Yury~E. García, Juan~G.
  Calvo, Hugo~G. Hidalgo, and Fabio Sanchez.
\newblock Assessing dengue fever risk in costa rica by using climate variables
  and machine learning techniques.
\newblock {\em PLOS Neglected Tropical Diseases}, 17(1):1--13, 01 2023.

\bibitem{GoodProbabilityAssessors}
Robert~L Winkler and Allan~H Murphy.
\newblock {“Good” Probability Assessors}.
\newblock {\em Journal of Applied Meteorology and Climatology}, 7(5):751--758,
  1968.

\bibitem{Gneiting2007a}
Tilmann Gneiting and Adrian~E Raftery.
\newblock {Strictly Proper Scoring Rules, Prediction, and Estimation}.
\newblock {\em Journal of the American Statistical Association},
  102(477):359--378, mar 2007.

\bibitem{racloz2012surveillance}
Vanessa Racloz, Rebecca Ramsey, Shilu Tong, and Wenbiao Hu.
\newblock Surveillance of dengue fever virus: a review of epidemiological
  models and early warning systems.
\newblock {\em PLoS neglected tropical diseases}, 6(5):e1648, 2012.

\bibitem{wen2006spatial}
Tzai-Hung Wen, Neal~H Lin, Chun-Hung Lin, Chwan-Chuen King, and Ming-Daw Su.
\newblock Spatial mapping of temporal risk characteristics to improve
  environmental health risk identification: a case study of a dengue epidemic
  in taiwan.
\newblock {\em Science of the Total Environment}, 367(2-3):631--640, 2006.

\end{thebibliography}

\pagebreak
\begin{center}
\textbf{\large Supplemental Materials: Bayesian spatio-temporal model with INLA for dengue fever risk prediction in Costa Rica}
\end{center}

\setcounter{equation}{0}
\setcounter{figure}{0}
\setcounter{table}{0}
\setcounter{page}{1}
\makeatletter
\renewcommand{\thefigure}{S\arabic{figure}}
\renewcommand{\thetable}{S\arabic{table}}

\appendix
\counterwithin{figure}{section}
\counterwithin{table}{section}

	\section{Naïve methods' predictive metrics of the testing period (from Juanuary to March 2021)}
	\label{app:naive_pred}
	
	\begin{table}[H]
		\centering
		\caption{\label{tab:naive_predict} Predictive metrics of testing data set of the naïve forecasting and negative binomial null model}
		\begin{threeparttable}
			\begin{tabular}[t]{lrrr}
				\toprule
				\multicolumn{1}{c}{Municipality} & \multicolumn{1}{c}{Naive method} & \multicolumn{2}{c}{Negative Binomial null model} \\
				\cline{2-2} \cline{3-4}
				& $NRMSE$ & $NRMSE$ & $NIS_{95}$\\
				\midrule
				Alajuela & 2.6506 & 1.4229 & 42.9843\\
				Alajuelita & 0.4765 & 2.1888 & 59.8866\\
				Atenas & 89.8765 & 7.3082 & 156.7569\\
				Cañas & 3.5234 & 2.9727 & 69.7955\\
				Carrillo & 0.8118 & 0.9751 & 26.7633\\
				Corredores & 0.3051 & 0.7459 & 20.5897\\
				Desamparados & 0.2598 & 22.3581 & 436.3609\\
				Esparza & 0.6795 & 0.9124 & 26.0958\\
				Garabito & 15.6450 & 9.0487 & 189.7400\\
				Golfito & 0.9871 & 3.6079 & 82.9118\\
				Guacimo & 3.6451 & 3.2932 & 28.8848\\
				La Cruz & 3.5519 & 17.4492 & 339.4951\\
				Liberia & 0.4927 & 16.3574 & 319.1370\\
				Limon & 3.4131 & 2.1836 & 23.9121\\
				Matina & 2.6768 & 0.8195 & 30.0298\\
				Montes de Oro & 41.8550 & 8.0778 & 160.1058\\
				Nicoya & 9.2749 & 17.0730 & 338.3457\\
				Orotina & 2.2117 & 0.9807 & 15.6226\\
				Osa & 1.2030 & 0.6291 & 13.9754\\
				Parrita & 15.7029 & 2.4511 & 54.8356\\
				Perez Zeledón & 1.5211 & 0.2745 & 7.9940\\
				Pococí & 1.0359 & 0.2630 & 10.7672\\
				Puntarenas & 0.7206 & 2.0884 & 52.3785\\
				Quepos & 10.8189 & 9.6902 & 192.1924\\
				San Jose & 0.0267 & 11.4795 & 237.4494\\
				Santa Ana & 0.4281 & 19.8108 & 383.3144\\
				SantaCruz & 2.3943 & 6.6282 & 144.8392\\
				Sarapiquí & 13.7461 & 0.0603 & 3.2807\\
				Siquirres & 1.8468 & 0.2815 & 9.1988\\
				Talamanca & 13.8805 & 14.4673 & 35.1501\\
				Turrialba & 1.4088 & 0.3812 & 14.9645\\
				Upala\tnote{1} & - & - & -\\
				\bottomrule
			\end{tabular}
			\begin{tablenotes}
				\item[1] $NRMSE$ and $NIS_{95}$ for Upala is not shown since the observed relative risks are zero.
			\end{tablenotes}
		\end{threeparttable}
	\end{table}

	\newpage

	\newpage
	
	\section{Dengue cases modelling and prediction}
	\label{app:prediction}
	
	\begin{figure}[H]
		\centering
		\includegraphics[scale=0.1]{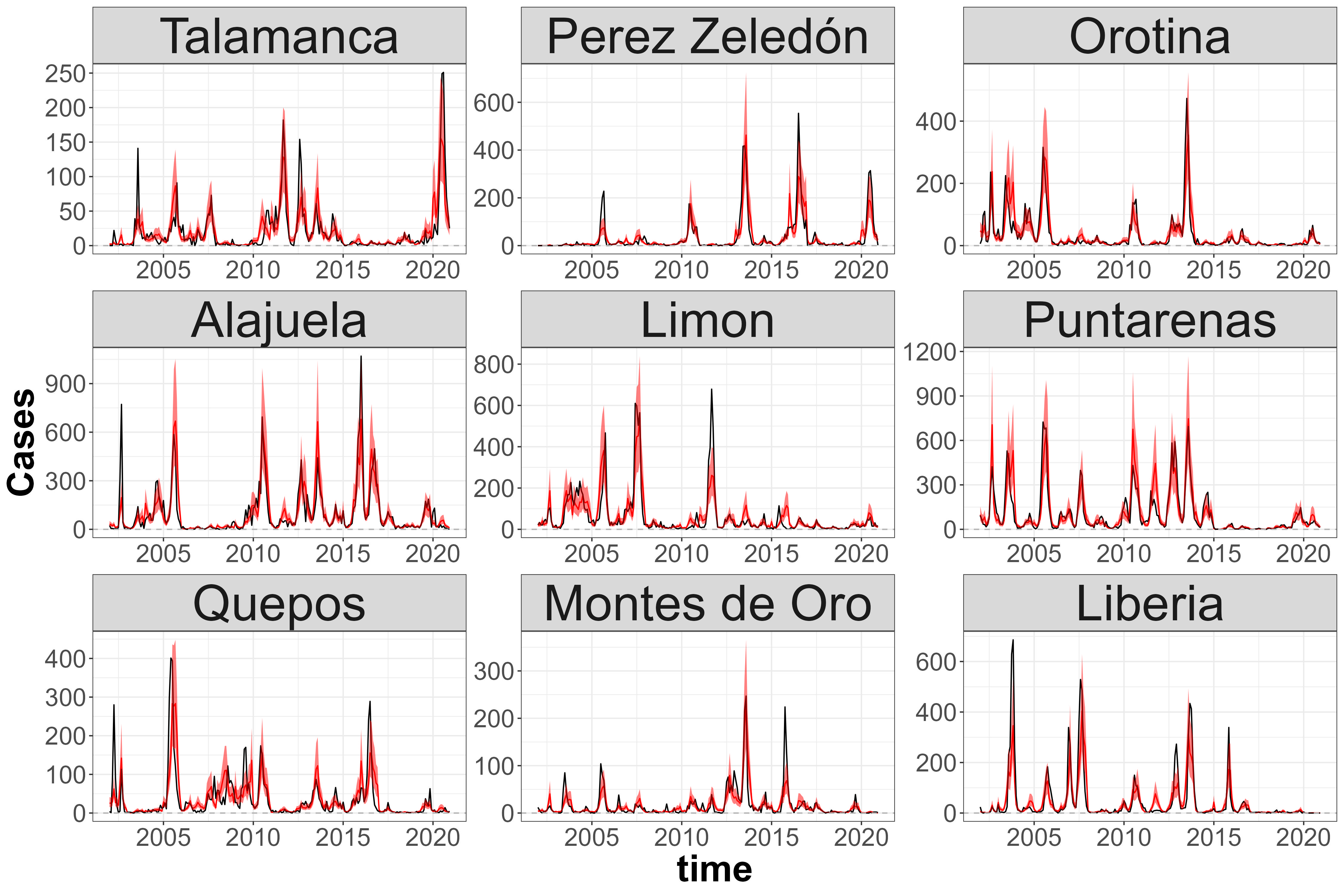}
		\caption{Observed (black) and 95\% posterior predictive dengue cases (red) over the training period. Upper six panels: best municipalities according to NIS metric. Lower three panels: worst municipalities according to NIS metric.}
		\label{fig:case.prediction.in}
	\end{figure}		
	
	\begin{figure}[H]
		\centering
		\includegraphics[scale=0.1]{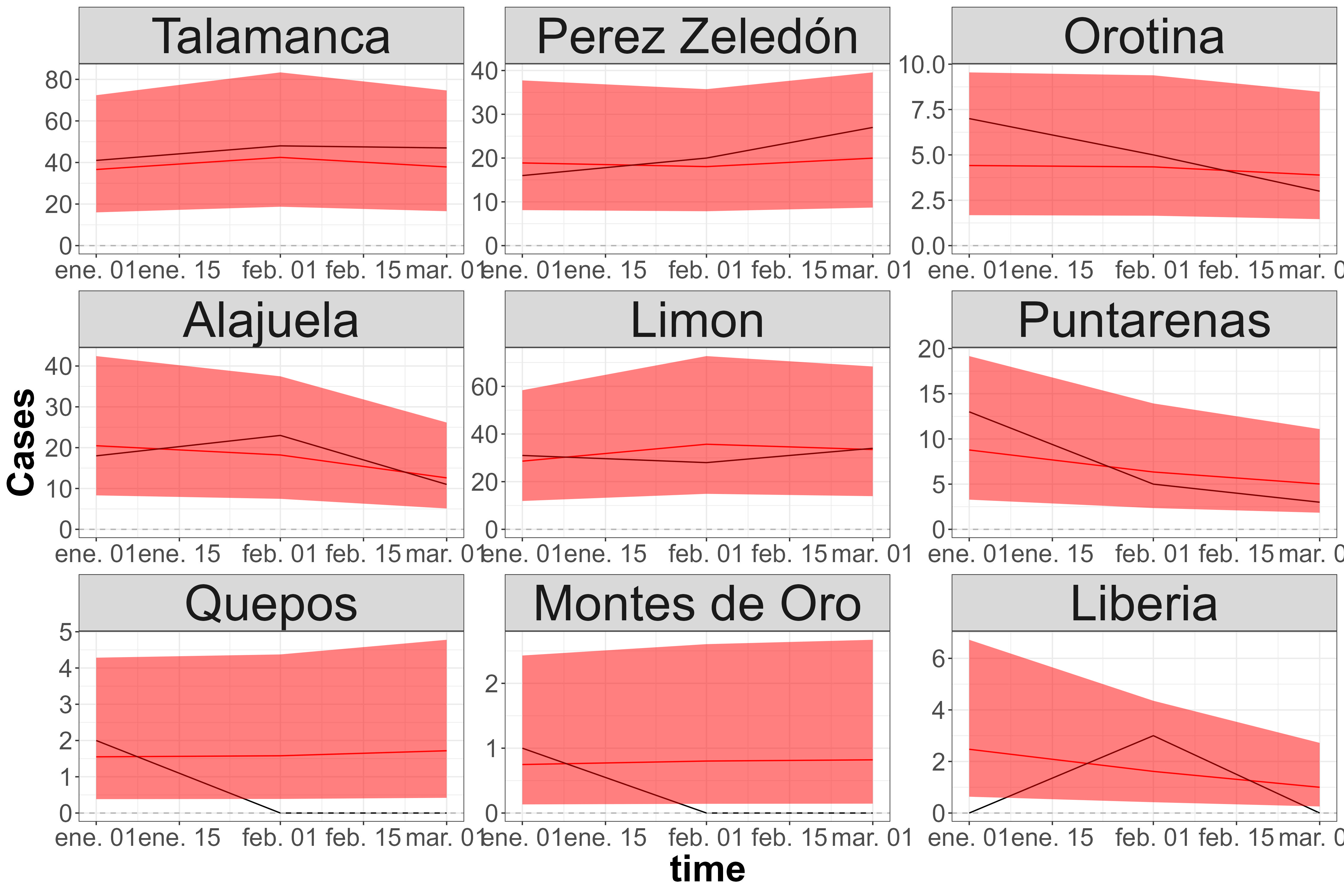}
		\caption{Observed (black) and 95\% posterior predictive dengue cases (red) over the testing period. Upper six panels: best municipalities according to NIS metric. Lower three panels: worst municipalities according to NIS metric.
		}
		\label{fig:case.prediction.out}
	\end{figure}		
	
\newpage

\section{Relative risk prediction maps in 2002, 2011 and 2020}

\label{app:prediction_map}

	\begin{figure}[H]
		\centering
		\includegraphics[scale=0.08]{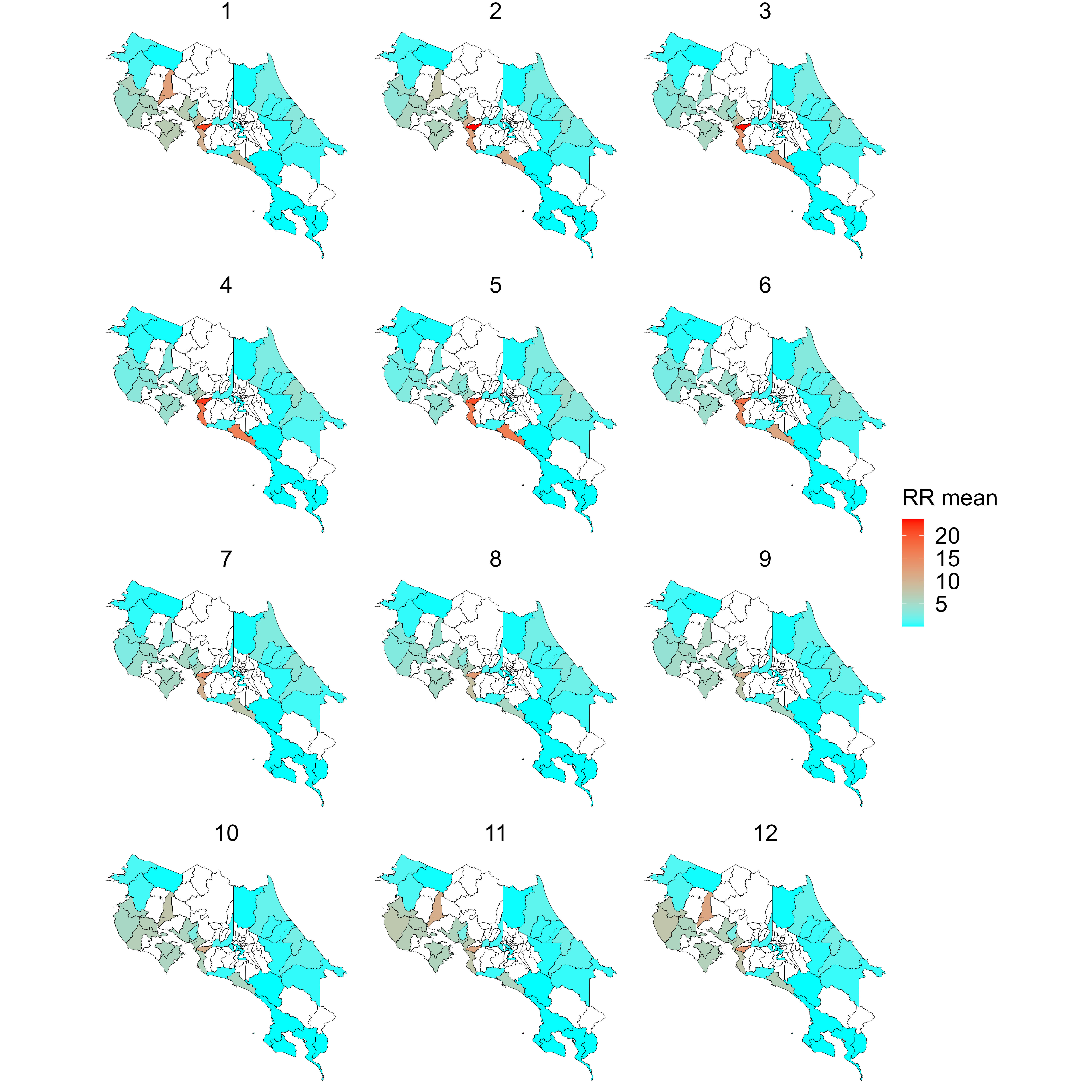}
		\caption{Posterior mean of relative risks from January to December 2002 for 81 municipalities in Costa Rica.}
		\label{fig:prediction_map2002}
	\end{figure}		
	
	\begin{figure}[H]
		\centering
		\includegraphics[scale=0.08]{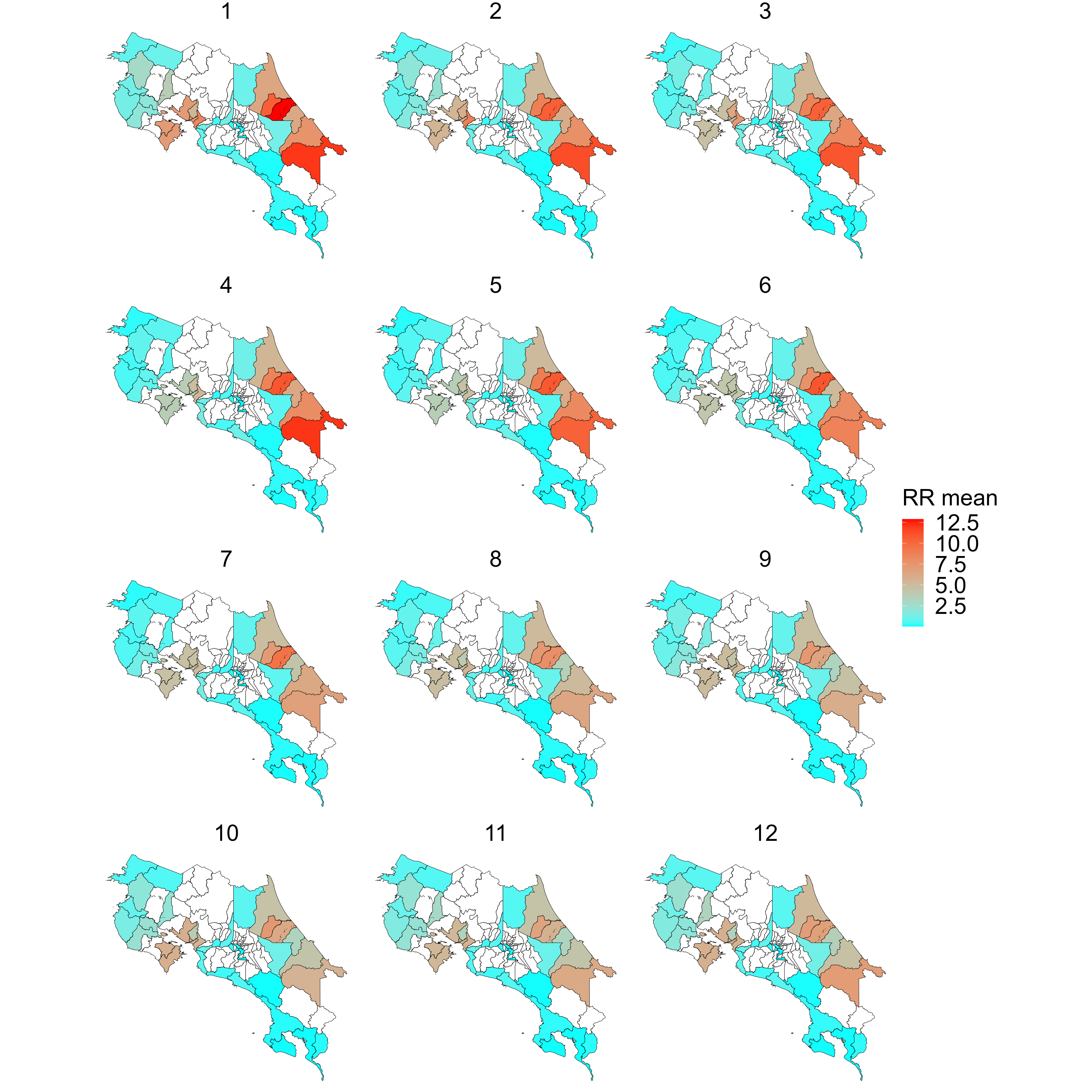}
		\caption{Posterior mean of relative risks from January to December 2011 for 81 municipalities in Costa Rica.}
		\label{fig:prediction_map2011}
	\end{figure}		
	
	\begin{figure}[H]
		\centering
		\includegraphics[scale=0.08]{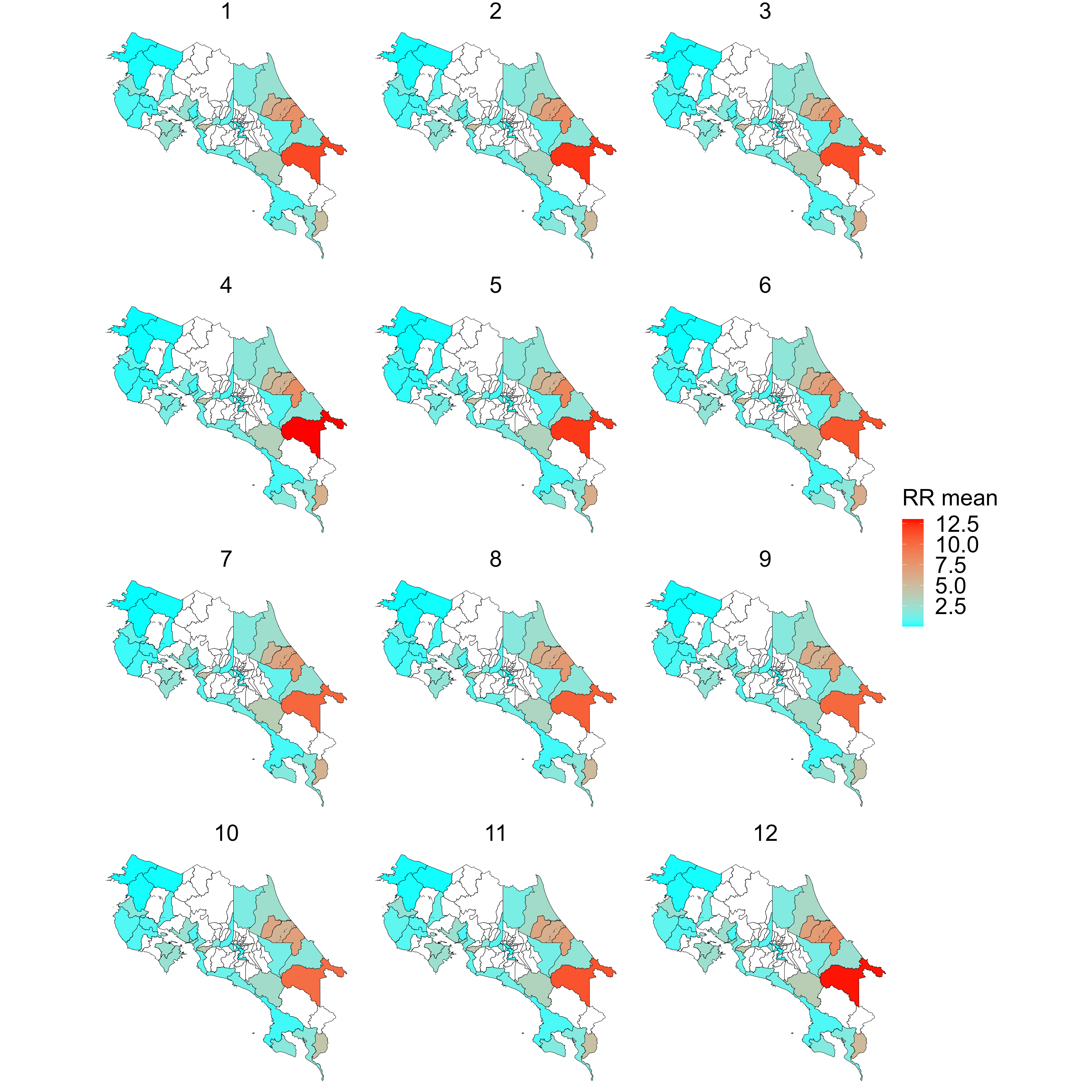}
		\caption{Posterior mean of relative risks from January to December 2020 for 81 municipalities in Costa Rica.}
		\label{fig:prediction_map2020}
	\end{figure}

\section{Absolute percentage error maps in 2002, 2011 and 2020}

	\begin{figure}[H]
		\centering
		\includegraphics[scale=0.08]{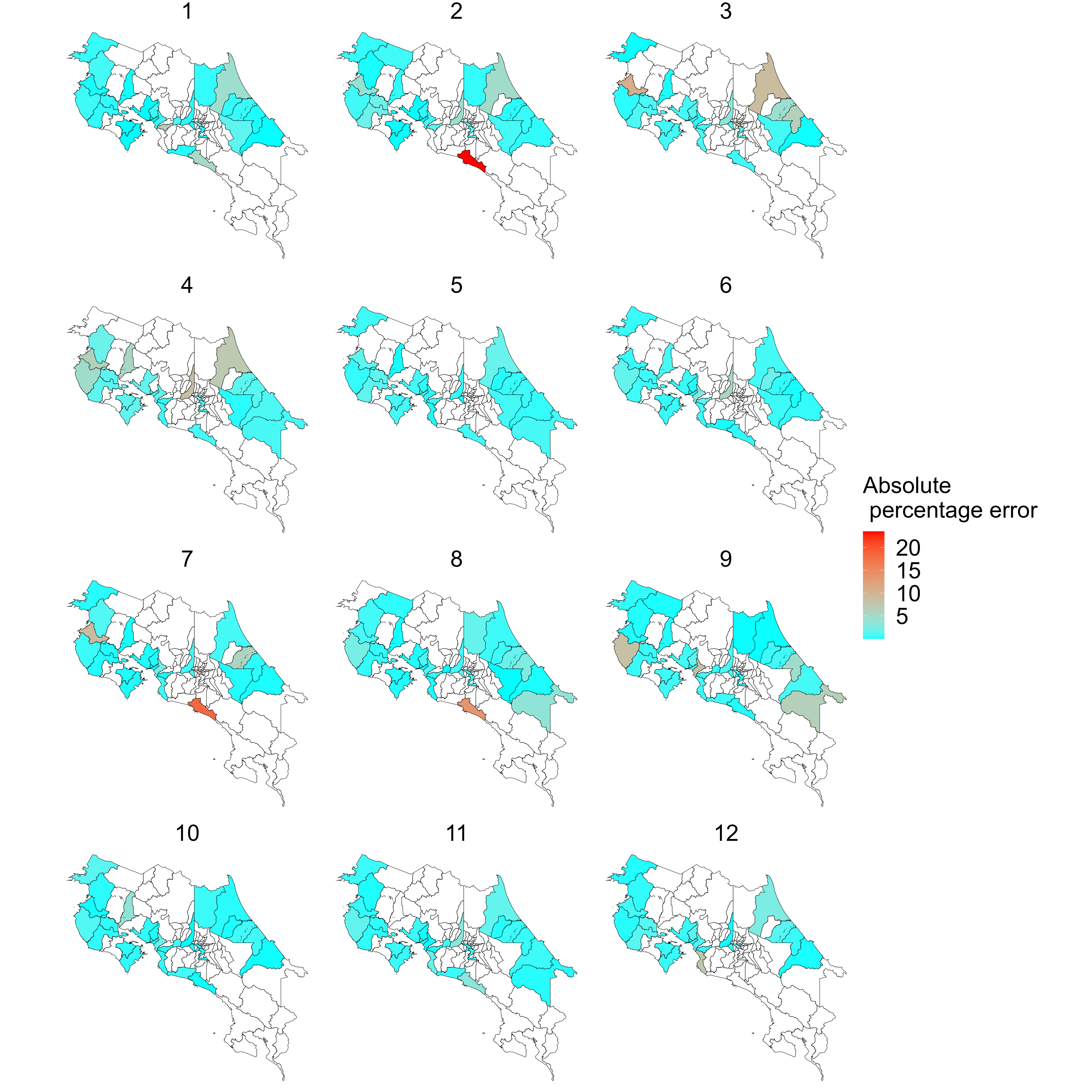}
		\caption{Absolute percentage error from January to December 2002 for 81 municipalities in Costa Rica.}
		\label{fig:dif_map2002}
	\end{figure}		
	
	\begin{figure}[H]
		\centering
		\includegraphics[scale=0.08]{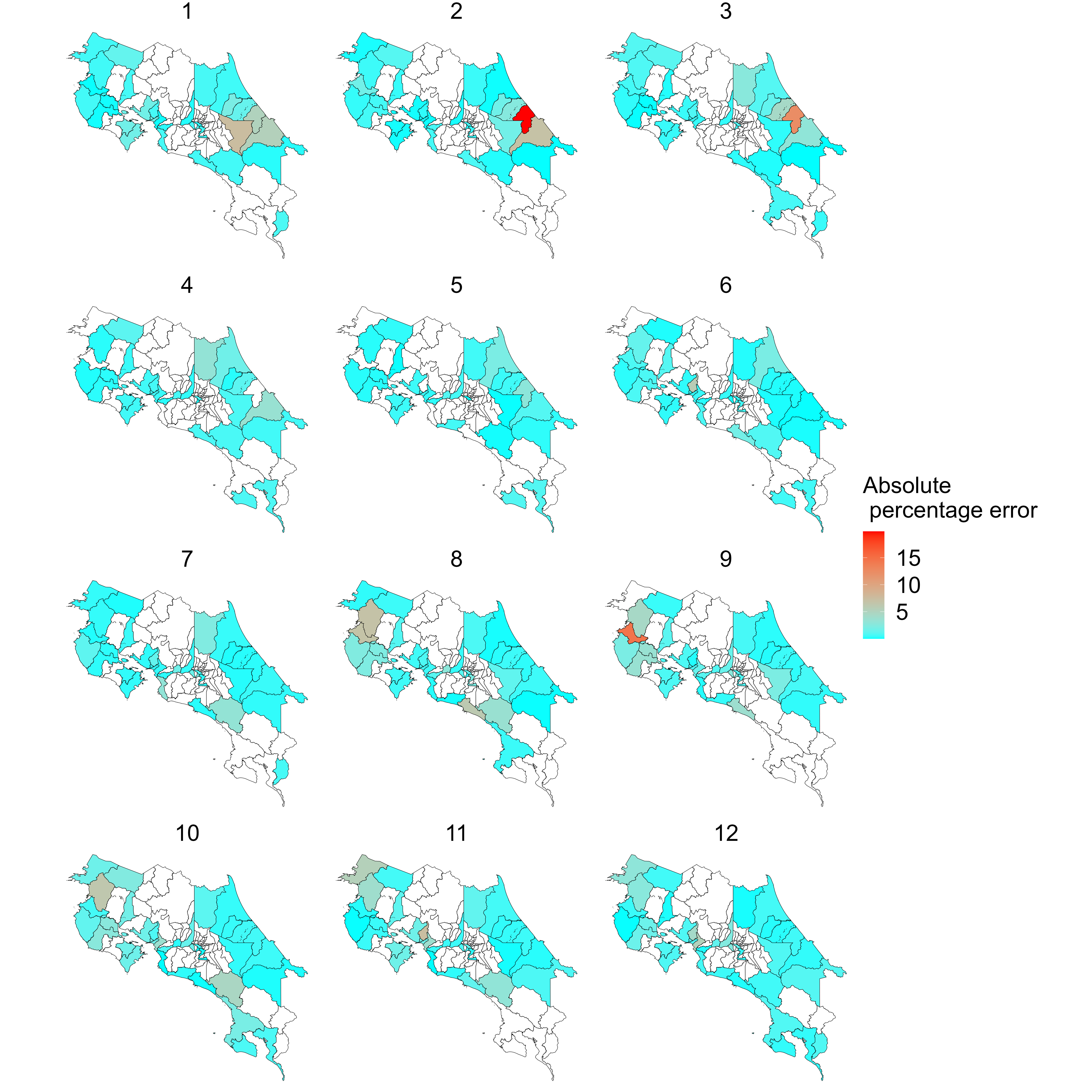}
		\caption{Absolute percentage error from January to December 2011 for 81 municipalities in Costa Rica.}
		\label{fig:dif_map2011}
	\end{figure}		
	
	\begin{figure}[H]
		\centering
		\includegraphics[scale=0.08]{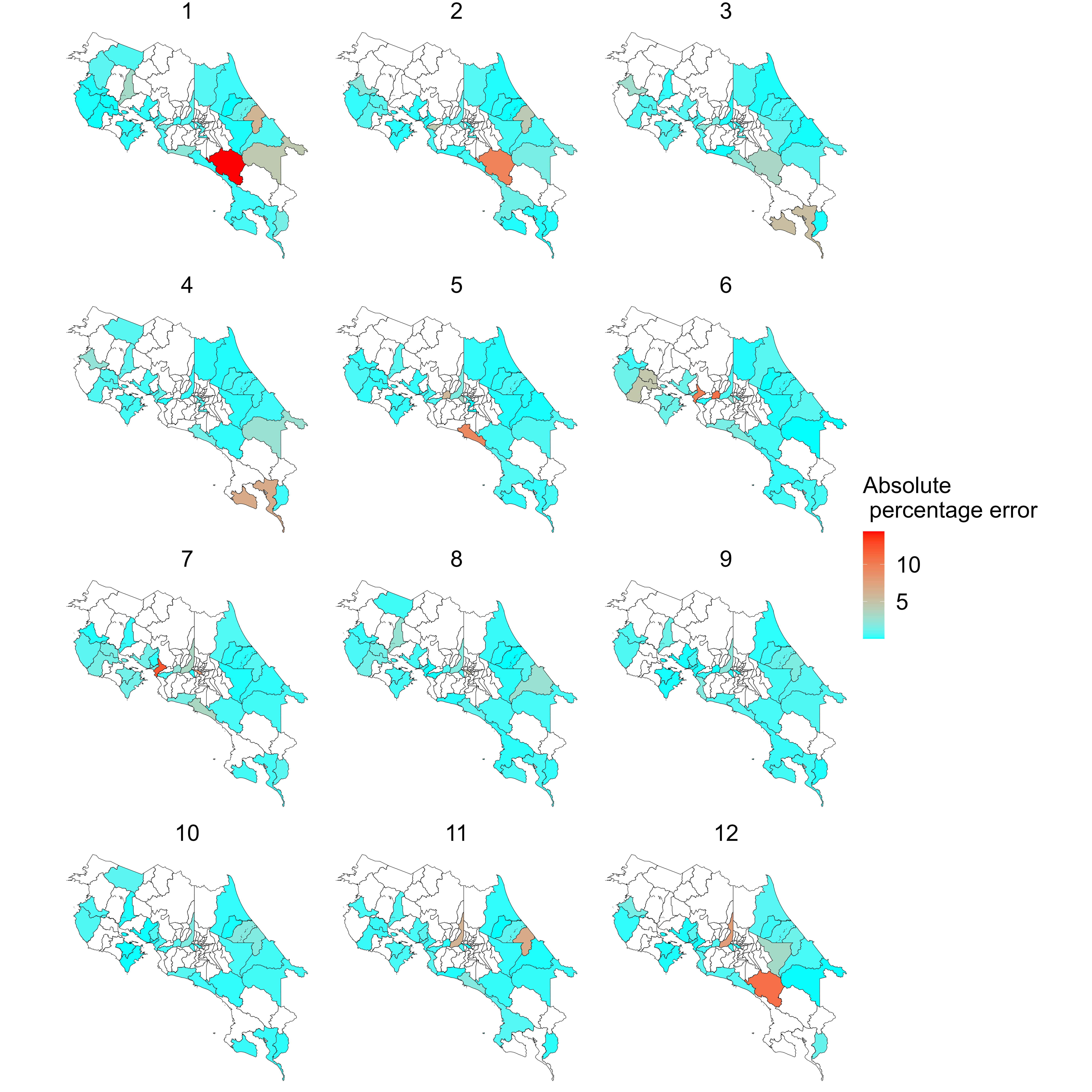}
		\caption{Absolute percentage error from January to December 2020 for 81 municipalities in Costa Rica.}
		\label{fig:dif_map2020}
	\end{figure}

\end{document}